\documentclass[journal]{IEEEtran}
\usepackage{cite}

\ifCLASSINFOpdf
   \usepackage[pdftex]{graphicx}
   \graphicspath{{../Simulation/}}
   \DeclareGraphicsExtensions{.pdf,.jpeg,.png}
\else
\fi
\usepackage{caption}
\usepackage{pgf}
\usepackage{tikz}
\usetikzlibrary{arrows,automata}
\usepackage[latin1]{inputenc}

\usepackage{amsmath,amsfonts,amsthm,bbm} 
\usepackage{amssymb}
\newtheorem{thm}{Theorem}
\newtheorem{corollary}{Corollary}
\newtheorem{proposition}{Proposition}
\newtheorem{lemma}{Lemma}

\usepackage{cleveref}

\ifCLASSOPTIONcompsoc
  \usepackage[caption=false,font=normalsize,labelfont=sf,textfont=sf]{subfig}
\else
  \usepackage[caption=false,font=footnotesize]{subfig}
\fi

\newcommand{\E}{\mathbb{E}}
\newcommand{\Prob}{\mathbb{P}}
\newcommand{\nn}{\nonumber\\}
\newcommand{\lp}{\left(}
\newcommand{\rp}{\right)}
\newcommand{\str}{\mathcal{U}_1}
\newcommand{\strr}{\mathcal{U}_2}

\newcommand{\XX}{X^{(2)}}
\newcommand{\Ss}{S^{(1)}}
\newcommand{\SsS}{S^{(2)}}

\begin{document}
%
\title{Content Based Status Updates}

\author{\IEEEauthorblockN{Elie Najm,}
\IEEEauthorblockA{LTHI, EPFL, Lausanne, Switzerland\\
}
\and 
\IEEEauthorblockN{Rajai Nasser,}
\IEEEauthorblockA{AUB, Beirut, Lebanon\\
}
\and
\IEEEauthorblockN{Emre Telatar,}
\IEEEauthorblockA{LTHI, EPFL, Lausanne, Switzerland\\
}
}


%


\maketitle

\begin{abstract}
Consider a stream of status updates generated by a source,
where each update is of one of two types: high priority or ordinary (low priority). These
updates are to be transmitted through a network to a monitor. However, the transmission policy of each packet depends on the
type of stream it belongs to. For the low priority stream, we
analyze and compare the performances of two transmission schemes: $(i)$ Ordinary updates are served in a First-Come-First-Served (FCFS)
fashion, whereas, in $(ii)$, the ordinary updates are transmitted according to an M/G/1/1 with preemption policy. In
both schemes, high priority updates are transmitted according to an M/G/1/1 with preemption policy and receive preferential treatment.
An arriving priority update discards and replaces any currently-in-service high
priority update, and preempts (with eventual resume for scheme $(i)$) any ordinary
update.  We model the arrival processes of the two kinds of updates, in both schemes, as
independent Poisson processes. 
For scheme $(i)$, we find the arrival and service rates
under which the system is stable and give closed-form expressions for 
average peak age and a lower bound on the average age of the ordinary stream.
For scheme $(ii)$, we derive closed-form expressions for the average age and average peak age of the high priority and low
priority streams. We finally show that, if the service time is exponentially distributed, the M/M/1/1 with preemption policy
leads to an average age of the low priority stream higher than the one achieved using the FCFS scheme. Therefore, the M/M//1/1
with preemption policy, when applied on the low priority stream of updates and in the presence of a higher priority scheme, is
not anymore the optimal transmission policy from an age point of view. 


\end{abstract}

\section{Introduction}
\label{sect:sect_intro}

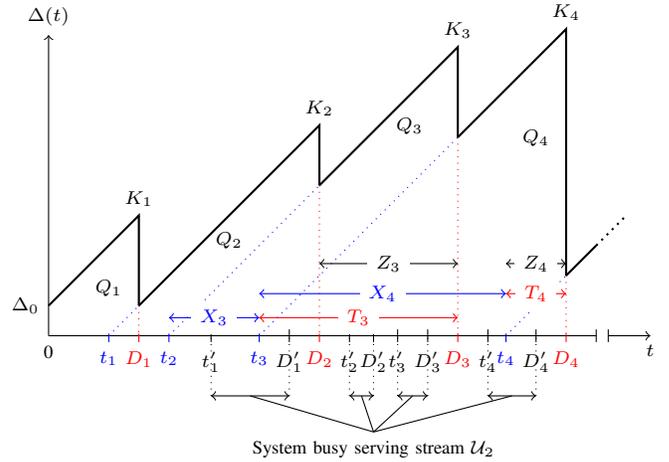
\begin{figure}[tp]
	\centering
	\begin{tikzpicture}[scale=0.8,font=\scriptsize]
		\draw (0,0) -- (9.1,0);
		\draw (9.1,3pt) -- (9.1,-3pt);
		\draw (9.3,3pt) -- (9.3,-3pt);
		\draw[->] (9.3,0) -- (10,0)node[anchor=north] {$t$};
		\draw	(0,0) node[anchor=north] {0};
		\draw[color=blue]	(1,1pt) -- (1,-3pt) node[anchor=north] {{$t_1$}};
		\draw[color=red]	(1.5,1pt) -- (1.5,-3pt) node[anchor=north] {$D_1$};
		
		\draw[color=blue]	(2,1pt) -- (2,-3pt) node[anchor=north] {$t_2$};
		\draw			(2.7,3pt) -- (2.7,-3pt) node[anchor=north] {$t'_1$};
		\draw[color=blue]	(3.5,1pt) -- (3.5,-3pt) node[anchor=north] {$t_3$};
		\draw			(4,3pt) -- (4,-3pt) node[anchor=north] {$D'_1$};
		\draw[color=red]	(4.5,1pt) -- (4.5,-3pt) node[anchor=north] {$D_2$};
		
		\draw			(5,3pt) -- (5,-3pt) node[anchor=north] {$t'_2$};
		\draw			(5.4,3pt) -- (5.4,-3pt) node[anchor=north] {$D'_2$};
		\draw			(5.8,3pt) -- (5.8,-3pt) node[anchor=north] {$t'_3$};
		\draw			(6.3,3pt) -- (6.3,-3pt) node[anchor=north] {$D'_3$};
		\draw[color=red]	(6.8,1pt) -- (6.8,-3pt) node[anchor=north] {$D_3$};
		
		\draw			(7.3,3pt) -- (7.3,-3pt) node[anchor=north] {$t'_4$};
		\draw[color=blue]	(7.6,1pt) -- (7.6,-3pt) node[anchor=north] {$t_4$};
		\draw			(8.1,3pt) -- (8.1,-3pt) node[anchor=north] {$D'_4$};
		\draw[color=red]	(8.6,1pt) -- (8.6,-3pt) node[anchor=north] {$D_4$};
					
		\draw[->] (0,0) -- (0,5) node[anchor=south] {$\Delta(t)$};
				
		\draw[thick] (0,0.5) -- (1.5,2) -- (1.5,0.5);
		\draw[dotted, color=blue] (1,0) -- (1.5,0.5);
		\draw[dotted, color=red]  (1.5,0.5) -- (1.5,0);
		\draw (0,0.5) node[anchor=east] {$\Delta_0$}; 
		\draw (1.5,2) node[anchor=south] {$K_1$};
		\draw (1,0.5) node[anchor=south] {$Q_1$};
		
		\draw[thick] (1.5,0.5) -- (4.5,3.5) -- (4.5,2.5);
		\draw[dotted, color=blue] (2,0) -- (4.5,2.5);
		\draw[dotted, color=red]  (4.5,2.5) -- (4.5,0);
		\draw (4.5,3.5) node[anchor=south] {$K_2$};
		\draw (3,1.3) node[anchor=south] {$Q_2$};
			
		\draw[thick] (4.5,2.5) -- (6.8,4.8) -- (6.8,3.3);
		\draw[dotted, color=blue] (3.5,0) -- (6.8,3.3);
		\draw[dotted, color=red]  (6.8,3.3) -- (6.8,0);
		\draw (6.8,4.8) node[anchor=south] {$K_3$};
		\draw (6,3.2) node[anchor=south] {$Q_3$};
		
		\draw[thick] (6.8,3.3) -- (8.6,5.1) -- (8.6,1);
		\draw[dotted, color=blue] (7.6,0) -- (8.6,1);
		\draw[dotted, color=red]  (8.6,1) -- (8.6,0);
		\draw (8.6,5.1) node[anchor=south] {$K_4$};
		\draw (8.1,3) node[anchor=south] {$Q_4$};

		\draw[thick] (8.6,1) -- (9.1,1.5);
		\draw[dotted, thick] (9.1,1.5)--(9.6,2);

%

		\draw[<->] (2.7,-1) -- (4,-1);
		\draw[<->] (5,-1) -- (5.4,-1);
		\draw[<->] (5.8,-1) -- (6.3,-1);
		\draw[<->] (7.3,-1) -- (8.1,-1);
		
		\draw[dotted] (2.7,-1) -- (2.7,0);
		\draw[dotted] (4,-1) -- (4,0);
		\draw[dotted] (5,-1) -- (5,0);
		\draw[dotted] (5.4,-1) -- (5.4,0);
		\draw[dotted] (5.8,-1) -- (5.8,0);
		\draw[dotted] (6.3,-1) -- (6.3,0);
		\draw[dotted] (7.3,-1) -- (7.3,0);
		\draw[dotted] (8.1,-1) -- (8.1,0);

		\draw (3.35,-1)--(5.4,-1.6);
		\draw (5.2,-1)--(5.4,-1.6);
		\draw (6.05,-1)--(5.4,-1.6);
		\draw (7.7,-1)--(5.4,-1.6) node[anchor=north] {System busy serving stream $\mathcal{U}_2$};
		
		\draw[<->, color=red] (3.5,0.3) -- (6.8,0.3);
		\draw[color=red] (5.15,0.3) node[anchor=center, fill=white] {$T_3$};
		\draw[<->, color=red] (7.6,0.7) -- (8.6,0.7);
		\draw[color=red] (8.1,0.7) node[anchor=center, fill=white] {$T_4$};

		\draw[<->, color=blue] (3.5,0.3) -- (2,0.3);
		\draw[color=blue] (2.75,0.3) node[anchor=center, fill=white] {$X_3$};
		\draw[<->, color=blue] (7.6,0.7) -- (3.5,0.7);
		\draw[color=blue] (5.55,0.7) node[anchor=center, fill=white] {$X_4$};

		\draw[<->] (4.5,1.2) -- (6.8,1.2);
		\draw		(5.65,1.2) node[anchor=center, fill=white] {$Z_3$};
		\draw[<->] (7.6,1.2) -- (8.6,1.2);
		\draw 		(8.1,1.2) node[anchor=center, fill=white] {$Z_4$};
	\end{tikzpicture}
	\caption{Variation of the instantaneous age of stream $\mathcal{U}_1$.}
	\label{fig:fig_ch5_fig1}
\end{figure}

While the classical notion of delay is a measure of how long a packet
spends in transit, the \lq Age of Information\rq\ \cite{KaulYatesGruteser-Globcom2011} is a
receiver-centric notion that measures how fresh the data is at the
receiver.  Specifically, with $u(t)$ denoting the generation time of the
last successfully received packet before time $t$, one defines $\Delta(t) =
t-u(t)$ as the instantaneous age of the information at the receiver at
time $t$.  One can then consider
\begin{equation}
	\label{eq:eq_age_definition}
	\Delta = \lim_{\tau\to\infty} \frac{1}{\tau}\int_0^\tau \Delta(t)\mathrm{d}t,
\end{equation} 
as the (time) average age.  Observe that $\Delta(t)$ increases linearly in
the intervals between packet receptions, and when a packet is received,
$\Delta(t)$ jumps down to the delay experienced by this packet.  This
results in a sawtooth sample path as in Fig.~\ref{fig:fig_ch5_fig1}.  In
\cite{KaulYatesGruteser-2012Infocom,2012CISS-KaulYatesGruteser,CostaCodreanuEphremides2014ISIT,
KamKompellaEphremides2013ISIT,NajmNasser-ISIT2016,YatesKaul-2012ISIT} the
properties of $\Delta$ were investigated under the assumption that the
packets are generated by a Poisson process, and various transmission
policies (M/M/1, M/M/$\infty$, gamma service time,\dots).

A related metric, called average peak age, was introduced in \cite{CostaCodreanuEphremides2014ISIT} as
the average of the value of the instantaneous age $\Delta(t)$ at times
just before its downward jumps.  In Fig.~\ref{fig:fig_ch5_fig1}, $K_j$ denotes the
instantaneous age just before the reception of the $j^{th}$ successfully
transmitted packet, and hence, the average peak age is given by
\begin{equation}
	\label{eq:eq_avg_peak_age_def}
	\Delta_{peak} = \lim_{N\to\infty}\frac{1}{N}\sum_{j=1}^N K_j.
\end{equation}

Yates et al., in \cite{YatesKaul-2016arxiv}, studied the average age when considering multiple sources sending update through one
queue. They computed the average age for three scenarios: all sources transmit according to an M/M/1 FCFS policy, all sources
transmit according to an M/M/1/1 with preemption policy and all sources transmit according to an M/M/1/1 with preemption in
waiting policy.
In the M/M/1/1 with preemption policy, if a newly generated update finds the system busy, the transmitter preempts the one
currently in service and starts sending the new packet. On the other hand, in the M/M/1/1 with preemption in waiting policy,
the system has a buffer of size 1 and if the generated
update finds the system busy, it replaces any waiting update in the buffer. 
In \cite{HuangModiano2015ISIT}, Huang et al. also consider multiple sources transmitting through a single queue but in this case
they assume a generally distributed service time. Moreover, they study  two scenarios: all sources transmit according to an
M/G/1 FCFS policy or all sources transmit according to an M/G/1/1 with blocking policy. For each one of these policies, the
authors give the expression of the average peak age of each source. 
In \cite{NajmTelatar2018}, Najm et al. also consider multiple sources sending through one queue. However, the authors derive
the closed-form expressions of the average age and average peak age when assuming an M/G/1/1 with preemption, with the service
time having a generic distribution. In their analysis, Najm et al. reintroduce the detour flow graph technique which we will
use extensively in this paper.

In this paper, we assume updates are generated according to a Poisson process with rate $\lambda$, and that updates belong to
two different streams where each stream $i$ is chosen independently with probability $p_i$, $i=1,2$. So we have
two independent Poisson streams with rates $\lambda_1=\lambda p_1$ and $\lambda_2 = \lambda p_2$. However, unlike
\cite{YatesKaul-2016arxiv,HuangModiano2015ISIT,NajmTelatar2018}, we assume a different transmission policy for each stream. The two
independent streams generated by the source can be used to model different types of content carried by the packets of each
stream. For example, if the source is a sensor,  one stream could carry emergency messages (fire alarm, high pressure, etc.)
and thus it needs to be always as fresh as possible while the other stream will carry regular updates and hence is not age
sensitive. Therefore, it stands to reason to transmit these two streams in a different manner. The paper is divided into
three parts: 
\begin{itemize}
	\item In the first part, we assume a different transmission policy for each stream. The regular stream will be
		transmitted according to a FCFS policy, whereas the high priority stream will be sent by preemption; packets of
		the high priority stream preempt all packets including packets of their own stream. We further assume that the
		service time requirements of the two streams are different. Although packets from both streams spend an
		exponential time in service, a packet of the regular stream is served at rate $\mu_1$, while  a packet of the
		high priority stream at rate $\mu_2$. This model was first presented by the authors of the current paper in
		\cite{NajmNasserTelatar-ISIT2018}. In this part, we will answer the following questions: What should the
		relation between $\lambda_1$, $\mu_1$, $\lambda_2$ and $\mu_2$ be for the system to be stable? How does each
		stream affect the average age of the other one? What are the ages of each stream? To answer these questions, we
		give a necessary and sufficient condition for the system stability and find the steady-state distribution of
		the underlying state-space. We also give closed-form expressions for both the average peak-age and a lower
		bound on the average age of the regular stream, and compare them to the average age of the high-priority stream.

	\item In the second part, we assume the same transmission policy for both streams. We use an M/G/1/1 with preemption
		scheme. However, we consider that a packet from the low-priority (or regular) stream is served according to a
		service-time distribution similar to that of the random variable $S_1$, whereas an update from the
		high-priority stream is served according to a service time distribution identical to that of the variable
		$S_2$. We denote by $f_{S_1}(t)$ and $f_{S_2}(t)$ the respective probability density functions (p.d.f) of these
		service times. In this part, we generalize part of the results presented in \cite{KaulYates-priorityISIT18},
		relative to the preemption policy, and derive closed-form expressions for the average age and average peak age
		for any type of service-time distribution. Kaul et al., in \cite{KaulYates-priorityISIT18}, address a similar
		problem. They consider multiple sources with different priorities with source $1$ given the highest priority
		and source $M$ the lowest priority. Two types of transmission schemes are investigated: $(i)$ an M/M/1/1 with
		preemption where any new packet from source $i$ preempts the packet currently in service if this update belongs
		to source $j$ with $j\geq i$, and $(ii)$ an M/M/1/2* where any new packet from source $i$ that finds the server
		busy would be placed in a buffer of size $1$. However, if the buffer is already occupied by an update from
		source $j$, $j\geq i$, then the waiting packet is dropped and replaced by the new one from source $i$. 

	\item In the third part, we compare, through simulations, the performance of the FCFS policy and that of the M/G/1/1
		with preemption on the age of the low priority stream. In this part, we show, through numerical results, that
		preemption is not the optimal transmission scheme to adopt, in the presence of higher priority streams,  even
		when the service time is exponential. This comes as a surprise since it was shown, by Bedewy et al. in
		\cite{BedewySunShroff-ISIT2016,BedewySunShroff17}, that the LCFS with preemption policy is optimal when we
		consider a single source generating updates according to a Poisson process and the service time at each hop is
		exponentially distributed. In fact, we observe that the FCFS policy achieves  a lower average age than the one
		achieved by the M/M/1/1 with preemption scheme. This means that if we are designing a system with a high and a
		low priority stream, and we have a choice between FCFS and M/G/1/1 with preemption as transmission schemes for
		the low priority stream, we should implement a FCFS transmission policy.  
\end{itemize}

This paper is structured as follows: In Section~\ref{sec:sec_ch5_system_model}, we start by defining the update
generation mechanism, common to both models and the different
variables needed in our study. In Section~\ref{sec:sec_ch5_fcfs_low_priority}, we study our first model and derive the stability condition of the system and its
stationary distribution. The closed-form expressions of the average peak-age and the lower bound on the average age of the
regular stream are computed in Section~\ref{sec:sec_ch5_ages_stream_1_2}. In Section~\ref{sec:sec_ch5_preempt_low_priority}, we
analyze the second model and compute the average age and average peak-age of both streams. Finally, in
Section~\ref{sec:sec_discussion}, we present some numerical results and show through an example that the FCFS policy
outperforms the M/M/1/1 with preemption, from the low priority stream point of view. 


\section{System Model}
\label{sec:sec_ch5_system_model}
We consider a sender that generates packets (or updates) according to a Poisson process of rate $\lambda$.  Each
packet, independently of the previous packets, is of type $1$ with probability $p_1$ and of type $2$ with probability
$p_2=1-p_1$.  We can thus see our sender as consisting of two sources generating two independent Poisson streams $\mathcal{U}_1$ and
$\mathcal{U}_2$ with rates
$\lambda_1=\lambda p_1$ and $\lambda_2=\lambda
p_2$ respectively, $\lambda=\lambda_1+\lambda_2$ (see \cite{ross}). As noted in the introduction, the different streams can 
be used to model packets of different types of content, for example, emergency messages, alerts, error messages, warnings, notices, etc.

We also assume that the updates are sent through a single server (or transmitter) to a monitor. The service times of 
packets from stream $\mathcal{U}_1$ are i.i.d according to $f_{S_1}(t)$, and those for stream
$\mathcal{U}_2$ are i.i.d according to $f_{S_2}(t)$. The difference in service rates between the two streams accounts for the possible difference in
compression, packet length, etc., between the two streams.
In \Cref{sec:sec_ch5_fcfs_low_priority}, the service time of each
packet is considered to be exponentially distributed, with rate $\mu_1$ for stream $\mathcal{U}_1$ and rate $\mu_2$ for stream
$\mathcal{U}_2$. However, in \Cref{sec:sec_ch5_preempt_low_priority} we keep the distributions general.


\section{FCFS for the Low-Priority Stream}
\label{sec:sec_ch5_fcfs_low_priority}
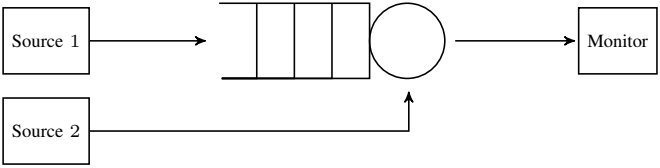
\begin{figure}
   \centering
   	\begin{tikzpicture}[	>=stealth',shorten >=1pt,auto,node distance=3.8cm,semithick,font=\scriptsize]
	\tikzstyle{every state}=[rectangle,fill=white,draw=black,text=black]
	\node[state] (A)					{Source $1$};
	\node[state] (B) [node distance=1.2cm,below of=A]	{Source $2$};

	\node (C) [right of=A]	{\begin{tikzpicture}[>=stealth',shorten >=1pt,auto,semithick]
			\draw (3,-1.2) --(5,-1.2)--(5,-2.2)--(3,-2.2) [right of=A];
			\draw (3.5,-1.2)--(3.5,-2.2)--(3,-2.2);
			\draw (4,-1.2)--(4,-2.2)--(3,-2.2);
			\draw (4.5,-1.2)--(4.5,-2.2)--(3,-2.2);
			\draw (5.5,-1.7) circle (0.5cm);\end{tikzpicture}};
	\node[state] (D) [right of=C]				{Monitor};
	
	\path[->] (A) edge node[] {} (C); 
	\draw[->] (B.east)-- ++(4,0)-| (C.328);
	\path[->] (C) edge node[] {} (D); 
		
	\end{tikzpicture}
	\caption{Diagram representing the model with FCFS for the low priority stream.}
	\label{fig:fig_ch5_fcfs_model}
\end{figure}
In this model, we constrain the transmitter so that all packets from stream $\mathcal{U}_1$ should be sent. Hence, the server
applies a FCFS policy on the packets from stream $\mathcal{U}_1$ with a buffer to save waiting updates. Whereas,
we assume that the information carried by stream $\mathcal{U}_2$ is more time sensitive (or has higher priority) hence we aim
to minimize its average age. To this end, the transmitter is permitted to perform packet management: In this case, we assume the server applies 
a preemption policy whenever a packet from $\mathcal{U}_2$ is generated. This means that if a newly generated packet from stream 
$\mathcal{U}_2$ finds the system busy (serving a packet from $\mathcal{U}_1$ or $\mathcal{U}_2$), the server preempts the
update currently in service and starts serving the new packet. On the one hand, if the preempted packet belongs to $\mathcal{U}_1$,
this packet is placed back at the head of the $\mathcal{U}_1$-buffer so that it can be served once the system is idle again. On
the other hand, if the preempted packet belongs to $\mathcal{U}_2$ then it is discarded. However, if a newly generated $\mathcal{U}_1$-packet
finds the system busy serving a $\mathcal{U}_2$-packet, it is placed in the buffer and served when the system becomes idle. This choice of policy for the
age sensitive stream is based on the conclusion reached in \cite{BedewySunShroff17}, that for exponentially distributed
packet transmission times, the M/M/1/1 with preemption policy is the optimal policy among causal policies.
Fig.~\ref{fig:fig_ch5_fcfs_model} gives a graphical representation of this model.

These ideas are illustrated in part in Fig.~\ref{fig:fig_ch5_fig1} which also shows the variation of the instantaneous age of stream
$\mathcal{U}_1$. In this plot, $t_i$ and $D_i$ refer to the generation and delivery times of the $i^{th}$ packet of stream
$\mathcal{U}_1$ while $t'_i$ and $D'_i$ are the start and end times of the $i^{th}$ period during which the system is busy
serving packets from stream $\mathcal{U}_2$ only. Notice that for stream $\mathcal{U}_1$ none of the generated packets is
discarded and all packets are received in the order of their generation.

\subsection{System Stability and Stationary Distribution}
\label{sec:sec_ch5_stab_system}

\begin{figure}[!t]
	\centering
\begin{tikzpicture}[>=stealth',shorten >=1pt,auto,node distance=2.4cm, semithick]
	\tikzstyle{every state}=[fill=red,draw=none,text=white]
	\node[state] (A)                     {$q_0$};
	\node[state] (B) [right of=A]  	     {$q_1$};
	\node[state] (C) [right of=B]  	     {$q_2$};
	\node[state] (D) [right of=C] 	     {$q_3$};
	\node[state,fill=white]	(E) at (8.5,0) {};
	\node[state] (A') [below of=A]       {$q'_1$};
	\node[state] (B') [below of=B]       {$q'_2$};
	\node[state] (C') [below of=C] 	     {$q'_3$};
	\node[state] (D') [below of=D] 	     {$q'_4$};
	\node[state,fill=white] (E') [below of=E]  {};
	
	\path [->] (A) edge [bend left]      node[ fill=white, anchor=center, pos=0.5] {$\lambda_1$} (B);
	\path [->] (A) edge [bend right]     node[ fill=white, anchor=center, pos=0.5] {$\lambda_2$} (A');
	
	\draw [->] (B) edge [bend left]      node[ fill=white, anchor=center, pos=0.5] {$\mu_1$} (A);
	\path [->] (B) edge [bend left]      node[ fill=white, anchor=center, pos=0.5] {$\lambda_1$} (C);
	\path [->] (B) edge [bend right]     node[ fill=white, anchor=center, pos=0.5] {$\lambda_2$} (B');
	
	\draw [->] (C) edge [bend left]      node[ fill=white, anchor=center, pos=0.5] {$\mu_1$} (B);
	\path [->] (C) edge [bend left]      node[ fill=white, anchor=center, pos=0.5] {$\lambda_1$} (D);
	\path [->] (C) edge [bend right]     node[ fill=white, anchor=center, pos=0.5] {$\lambda_2$} (C');
	
	\draw [->] (D) edge [bend left]      node[ fill=white, anchor=center, pos=0.5] {$\mu_1$} (C);
	\path [-]  (D) edge [bend left]      node {} (E);
	\path [->] (D) edge [bend right]     node[ fill=white, anchor=center, pos=0.5] {$\lambda_2$} (D');
	\path [->] (E) edge [bend left]      node {} (D);

	\path [->] (A') edge [bend left]     node[ fill=white, anchor=center, pos=0.5] {$\lambda_1$} (B');
	\path [->] (A') edge [bend right]    node[ fill=white, anchor=center, pos=0.5] {$\mu_2$} (A);
	\path [->] (B') edge [bend left]     node[ fill=white, anchor=center, pos=0.5] {$\lambda_1$} (C');
	\path [->] (B') edge [bend right]    node[ fill=white, anchor=center, pos=0.5] {$\mu_2$} (B);
	\path [->] (C') edge [bend left]     node[ fill=white, anchor=center, pos=0.5] {$\lambda_1$} (D');
	\path [->] (C') edge [bend right]    node[ fill=white, anchor=center, pos=0.5] {$\mu_2$} (C);
	\path [->] (D') edge [bend right]    node[ fill=white, anchor=center, pos=0.5] {$\mu_2$} (D);
	\path [-]  (D') edge [bend left]     node {} (E');
\end{tikzpicture}
\caption{Markov chain governing the number of packets in the system.}
\label{fig:fig_ch5_sys_mc}
\end{figure}
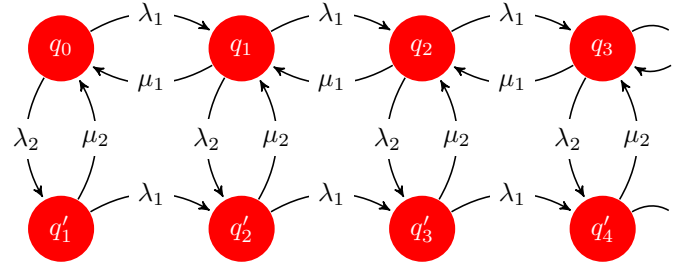

The fact that we seek to receive all of stream $\mathcal{U}_1$ updates and that stream $\mathcal{U}_2$ has a higher priority and
preempts stream $\mathcal{U}_1$ might lead to an unstable system. In order to derive the necessary and sufficient condition for
the stability of the system, we study the Markov chain of the number of packets in the system (in service and waiting) shown in
Fig.~\ref{fig:fig_ch5_sys_mc}. In this chain, $q_0$ is the idle state where the system is completely empty. States $q_i$, $i>0$, in
the upper row refer to states where the queue is serving a packet from stream $\mathcal{U}_1$, whereas states $q'_i$, $i>0$, in 
the row below correspond to the queue serving a packet from stream $\mathcal{U}_2$. In both cases, there are $i-1$ stream
$\mathcal{U}_1$ updates waiting in the buffer.

The system leaves state $q_0$ at rate $\lambda_1$ to state $q_1$ when a packet from stream $\mathcal{U}_1$ is generated first
and it leaves $q_0$ at rate $\lambda_2$ to state $q'_1$ when a packet from stream $\mathcal{U}_2$ is generated first. However,
when the system enters state $q_i$, $i>0$, three exponential clocks start: $(i)$ a clock with rate $\mu_1$, which corresponds to the
service time of the stream $\mathcal{U}_1$ packet being served, $(ii)$ a clock with rate $\lambda_1$, which corresponds to the
generation time of stream $\mathcal{U}_1$ packets and $(iii)$ a clock with rate $\lambda_2$, which corresponds to the generation time of
stream $\mathcal{U}_2$ packets. If the $\mu_1$-clock ticks first, the system goes to state $q_{i-1}$: This means that the
current stream $\mathcal{U}_1$ packet was delivered and the queue begins the service of the next one in the buffer (if there is
any). However, if the $\lambda_1$-clock ticks first, a new stream $\mathcal{U}_1$ update is generated and added to the
buffer, hence the system goes to state $q_{i+1}$. Whereas, if the $\lambda_2$-clock ticks first, the system
preempts the packet currently in service and places it back at the head of the buffer and starts the service of the newly
generated stream $\mathcal{U}_2$ update. Thus the system goes to state $q'_{i+1}$. When the system enters a state $q'_i$,
$i>0$, two exponential clocks start: the clock with rate $\lambda_1$ and a clock with rate $\mu_2$, which corresponds to the
service time of a stream $\mathcal{U}_2$ packet. If the $\lambda_1$-clock ticks first, the newly generated stream
$\mathcal{U}_1$ packet is placed in the buffer and the stream $\mathcal{U}_2$  update is continued to be served. Hence the
system goes to state $q'_{i+1}$. However, if the $\mu_2$-clock ticks first, the stream $\mathcal{U}_2$ packet has finished
service and the system starts serving the first stream $\mathcal{U}_1$ packet in the buffer (if there is any). Hence the system
goes to state $q_{i-1}$.

This next theorem gives the necessary and sufficient condition for the above system to be stable, as well as its stationary
distribution.
\begin{thm}
	\label{thm:thm_ch5_stat_dist}
	The system described in Section~\ref{sec:sec_ch5_fcfs_low_priority} is stable, i.e. the average number of packets in the queue
	is finite, if and only if 
	\begin{equation}
		\label{eq:eq_ch5_stab_condition}
		\mu_1 > \lambda_1\lp 1+\frac{\lambda_2}{\mu_2}\rp.
	\end{equation}
	In this case the Markov chain shown in Fig.~\ref{fig:fig_ch5_sys_mc} has a stationary distribution $\Pi=[\pi_0,
	\pi_1,\dots,\pi_i,\dots,\pi'_1,\dots,\pi'_i,\dots]$, where $\pi_i$ denotes the stationary probability of state
	$q_i$, $i\geq 0$, and $\pi'_i$ denotes the stationary probability of state $q'_i$, $i>0$. This stationary
	distribution is described by the following system of equations,
	\begin{align}
		\pi_0 & = \frac{\mu_2}{\mu_2+\lambda_2}-\frac{\lambda_1}{\mu_1}, \label{eq:eq_ch5_po}\\
		\begin{bmatrix}
			\pi_i\\
			\pi'_i
		\end{bmatrix} &= \begin{bmatrix}\mathbf{\large{0}} & \mathbf{I}_2\end{bmatrix}\mathbf{H}^i\begin{bmatrix}
			\frac{\lambda}{\mu_1}-\frac{\mu_2\lambda_2}{\mu_1\lp\lambda_1+\mu_2\rp}\\
			\frac{\lambda_2}{\lambda_1+\mu_2}\\
			1\\
			0
		\end{bmatrix}\pi_0,\ i\geq1 \label{eq:eq_ch5_stat_prob}
	\end{align}
	where $\lambda=\lambda_1+\lambda_2$, $ 
		\mathbf{H} = \begin{bmatrix}
			\mathbf{C} & \mathbf{D}\\
			\mathbf{I}_2 & \mathbf{\large{0}}
		\end{bmatrix}$,
	\begin{align*}
		&\mathbf{C} = \begin{bmatrix}
			1+\frac{\lambda}{\mu_1}-\frac{\mu_2\lambda_2}{\mu_1\lp\mu_2+\lambda_1\rp}  &
			-\frac{\mu_2\lambda_1}{\mu_1\lp\mu_2+\lambda_1\rp}\\
			\frac{\lambda_2}{\mu_2+\lambda_1}  & \frac{\lambda_1}{\mu_2+\lambda_1}
		\end{bmatrix}, \mathbf{D} = \begin{bmatrix}
			-\frac{\lambda_1}{\mu_1} & 0\\
			0	& 	0
		\end{bmatrix}.
	\end{align*}
	$\mathbf{I}_2$ is the $2\times2$ identity matrix and $\mathbf{\large{0}}$ is the $2\times2$ zero matrix.
\end{thm}

\begin{corollary}
\label{cor:cor_ch5_exp_value}
If we define $N(t)$ to be the number of stream $\mathcal{U}_1$ packets in the system at time $t$, then its moment generating
function is $\phi_{N(t)}$
\begin{equation}
	\label{eq:eq_ch5_mgf_stream_1}
	\resizebox{0.48\textwidth}{!}{$\phi_{N(t)}(s) = \pi_0\lp\frac{\mu_1\lp\lambda_1+\lambda_2+\mu_2-\lambda_1
	e^s\rp}{\mu_1\mu_2+\mu_1\lambda_1-e^s\lp\lambda_1^2+\lambda_1\lambda_2+\lambda_1\mu_1+\lambda_1\mu_2\rp+\lambda_1^2e^{2s}}\rp$},
\end{equation}
where $\pi_0$ is given by \eqref{eq:eq_ch5_po}. Particularly, the expected value of $N(t)$ is
\begin{equation}
\label{eq:eq_ch5_exp_stream_1}
\E\lp N(t)\rp = \frac{\lambda_1\lp 2\lambda_2\mu_2+\lambda_2\mu_1+\lambda_2^2+\mu_2^2\rp}{\lp \mu_2+\lambda_2\rp\lp
\mu_1\mu_2-\lambda_1\lp \mu_2+\lambda_2\rp\rp}.
\end{equation}
\end{corollary}
\begin{proof}
	The distribution given by \eqref{eq:eq_ch5_po} and \eqref{eq:eq_ch5_stat_prob} satisfy the detailed balance equations of the
	Markov chain shown in Fig.~\ref{fig:fig_ch5_sys_mc}.
	Moreover, \eqref{eq:eq_ch5_stab_condition} is the condition needed to have $\pi_0>0$. As for the expression for
	$\phi_{N(t)}(s)$, it is a
	consequence of \eqref{eq:eq_ch5_po} and \eqref{eq:eq_ch5_stat_prob}. The appendix in
	\Cref{subsec:subsec_ch5_proof_stationary_dist} and
	\Cref{appendix:appendix_proof_cor_ch5_exp_value} presents a full technical
version of the proof for Theorem~\ref{thm:thm_ch5_stat_dist} and Corollary~\ref{cor:cor_ch5_exp_value}. 
\end{proof}

\subsection{Interpretation of the stability condition} 
\label{subsec:subsec_interpreting_stability}
The condition in \eqref{eq:eq_ch5_stab_condition} can be interpreted in two equivalent ways:
\begin{enumerate}
	\item For an M/M/1 system with one source and an update rate $\lambda_1$ and service rate $\mu_1$, we need
		$\mu_1>\lambda_1$ for the system to be stable. However, in the case of stream $\mathcal{U}_1$ we need to
		compensate for the amount of time the second stream occupies the system. This explains the additional
		$\frac{\lambda_1\lambda_2}{\mu_2}$ term in \eqref{eq:eq_ch5_stab_condition} compared to an M/M/1 system.
	\item  Define  the map $f$ from the state-space of the chain as $f(s) = 0$ if $s$ is in  $\{q_0,q_1,\dots\}$ and
		$f(s)=1$ if $s\in\{q_1',q_2',\dots\}$.  For each $s$ and $s'$ for  which $f(s)=0$ and $f(s')=1$ the transition
		rate from $s$ to $s'$ is the same  $(\lambda_2)$ and similarly for $s$ and $s'$ with $f(s)=1$, $f(s')=0$,
		$(\mu_2)$.  Consequently $F(t)=f(s(t))$, with $s(t)$ being the state at time $t$, is Markov (which would not be
		the case for an  arbitrary $f$), and it is easily seen that $F(t)=0$ a fraction
		$\phi_0=\mu_2/(\lambda_2+\mu_2)$ amount of time, $F(t)=1$ a fraction  $\phi_1=\lambda_2/(\lambda_2+\mu_2)$
		amount of time.  This means that $\phi_0$ is the fraction of time spent by the system serving $\str$ packets or
		being idle, and $\phi_1$ is the fraction of time the system spends serving $\strr$ packets. The Markov chain
		representing the process $F(t)$ is given by Fig.~\ref{fig:fig_ch5_collapsed_sys_mc}. Thus, while the
		Markov  chain in Fig.~\ref{fig:fig_ch5_sys_mc} moves right at rate $\lambda_1$, it moves left at a
		rate  $\mu_1\phi_0$.  The system is stable only if the rate of moving left is larger than the rate of moving
		right; which gives the condition \eqref{eq:eq_ch5_stab_condition}.
\end{enumerate}
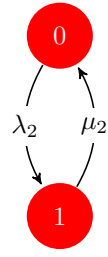
\begin{figure}[!t]
	\centering
\begin{tikzpicture}[>=stealth',shorten >=1pt,auto,node distance=2.4cm, semithick]
	\tikzstyle{every state}=[fill=red,draw=none,text=white]
	\node[state] (A)                     {$0$};
	\node[state] (A') [below of=A]       {$1$};
	
	\path [->] (A) edge [bend right]     node[ fill=white, anchor=center, pos=0.5] {$\lambda_2$} (A');
	
	\path [->] (A') edge [bend right]    node[ fill=white, anchor=center, pos=0.5] {$\mu_2$} (A);
\end{tikzpicture}
\caption{Markov chain representing whether the system is serving $\strr$ packets (state $1$) or not (state $0$).}
\label{fig:fig_ch5_collapsed_sys_mc}
\end{figure}



\subsection{Ages of Streams\ $\mathcal{U}_1$ and\ $\mathcal{U}_2$}
\label{sec:sec_ch5_ages_stream_1_2}

\subsubsection{Preliminaries}
\label{subsec:subsec_ch5_preliminaries}
In this section, unless stated otherwise, all random variables correspond to stream $\mathcal{U}_1$. We also follow the
convention where a random variable $U$ with no subscript corresponds to the steady-state version of $U_j$ that refers to the
random variable relative to the $j^{th}$ received packet from stream $\mathcal{U}_1$. To differentiate between streams, we will
use superscripts, which means that $U^{(i)}$ corresponds to the steady-state variable $U$ relative to stream $\mathcal{U}_i$ , $i=1,2$. 

In addition to this, we adopt the following notation:  
\begin{itemize}
	\item  $X^{(i)}$ is the interarrival time between two consecutive generated updates from stream
$\mathcal{U}_i$, so $f_{X^{(i)}}(x) = \lambda_i e^{-\lambda_i x}$, $i=1,2$ 
	\item  $S^{(i)}$ is the service time random variable of stream $\mathcal{U}_i$ updates, so $f_{S^{(i)}}(t) = \mu_i
		e^{-\mu_i t}$, $i=1,2$, 
	\item  $T_j$ is the system time, or the total time spent by the $j^{th}$ stream-$\mathcal{U}_1$ update in the queue (sum of
its waiting time and its service time). 
\end{itemize}
In our model, we assume the service time of the updates from the different streams to be independent of the interarrival time
between consecutive packets (belonging to the same stream or not). 

\subsubsection{Analysis of the System}
\label{subsubsec:subsubsec_ch5_analysis}
Given the aforementioned description of the model, we can define for each 
$\mathcal{U}_1$  packet $j$ a \lq\lq virtual\rq\rq\ service
time $Z_j$ that could be different from its \lq\lq physical\rq\rq\ service time $S^{(1)}_j$. We define the \lq\lq
virtual\rq\rq\ service time $Z_j$ as follows: 
\begin{equation}
	\label{eq:eq_ch5_Z}
	Z_j = D_j - \max(D_{j-1},t_j),
\end{equation}
where $D_j$ is the delivery time of the $j^{th}$ packet and
$t_j$ is its generation time. Fig.~\ref{fig:fig_ch5_fig1} shows the \lq\lq virtual\rq\rq\ service time for packets $3$ and $4$.

For stream $\mathcal{U}_1$, given that the average age calculations seem to be intractable, we compute its average peak
age and give a lower bound on its average age. To this end, we first study the steady state \lq\lq virtual\rq\rq\ service time
$Z$.

We define the event $$\Psi_j = \left\{\text{packet $j$
finds the system in state $q'_1$}\right\}$$ and its complement $\overline{\Psi_j}$. Then, we need the following lemmas.
\begin{lemma}
	\label{lemma:lemma_ch5_service_time_1}
		Let $Y_j$ be the \lq\lq virtual\rq\rq\ service time of packet $j$ given that this packet does not
			find the system in state $q'_1$, i.e. $\Prob\lp Y_j>t\rp=\Prob\lp Z_j>t|\overline{\Psi_j}\rp$. Then,
			in steady state,
			\begin{equation}
				\label{eq:eq_ch5_mgf_Y}
				\phi_Y(s) = \E\lp e^{sY}\rp= \frac{\mu_1(\mu_2-s)}{s^2-s(\mu_2+\mu_1+\lambda_2)+\mu_1\mu_2}.
			\end{equation}
		 Similarly, let $Y'_j$ be the \lq\lq virtual\rq\rq\ service time of packet $j$ given that this packet 
			finds the system in state $q'_1$, i.e. $\Prob\lp Y'_j>t\rp=\Prob\lp Z_j>t|\Psi_j\rp$. Then, in steady
			state,
			\begin{equation}
				\label{eq:eq_ch5_mgf_Y'}
				\phi_{Y'}(s) = \E\lp e^{sY'}\rp= \frac{\mu_1\mu_2}{s^2-s(\mu_2+\mu_1+\lambda_2)+\mu_1\mu_2}.
			\end{equation}
\end{lemma}
\begin{proof}
	This proof is based on the detour flow graph (or signal flow graph) method. An overview of this method as well as the
	complete proof are presented in Section~\ref{appendix:appendix_proof_lemma_ch5_service_time_1}.
\end{proof}

\subsubsection{Average Peak-Age of Stream $\mathcal{U}_1$}
\label{subsec:subsec_ch5_age_stream_2}
It is worth noting that the system under consideration cannot be seen as an M/G/1 queue with service time distributed as $Z$,
because the \lq\lq virtual\rq\rq\ service times of different packets are correlated. Indeed, if we know that the \lq\lq
virtual\rq\rq\ service time of packet $j$, $Z_j$, is big, then with very high probability the $(j+1)^{th}$ packet will be
generated during the service of the $j^{th}$ packet. Hence, with high probability, $Z_{j+1}$ will be distributed as $Y$. 
Whereas, if $Z_j$ is small, then there is a non-negligible probability with which the $(j+1)^{th}$ packet will find the
system serving stream $\mathcal{U}_2$. Hence, $Z_{j+1}$ will be distributed as $Y'$. 
\begin{thm}
	\label{thm:thm_ch5_avg_peak_age}
	The average peak age of stream $\mathcal{U}_1$ is given by
	\begin{equation}
		\label{eq:eq_ch5_avg_peak_age_1}
		\Delta_{peak,1} = \frac{1}{\lambda_1}+\frac{ 2\lambda_2\mu_2+\lambda_2\mu_1+\lambda_2^2+\mu_2^2}{\lp \mu_2+\lambda_2\rp\lp
\mu_1\mu_2-\lambda_1\lp \mu_2+\lambda_2\rp\rp}.
	\end{equation}
\end{thm}
\begin{proof}
	As we can deduce from Fig.~\ref{fig:fig_ch5_fig1}, the $j^{th}$ peak $K_j=X_j^{(1)}+T_j$ where $X_j^{(1)}$ is the $j^{th}$ interarrival time
	for stream $\mathcal{U}_1$ and $T_j$ is the system time of the $j^{th}$ stream $\mathcal{U}_1$ update. At steady state, we get
	$\Delta_{peak,1}=\E\lp K\rp = \E\lp X^{(1)}\rp + \E\lp T\rp$. From Little's law we know that $\E\lp T\rp = \E\lp
	N(t)\rp\E\lp X^{(1)}\rp$, with the expected number of stream $\mathcal{U}_1$ packets $\E\lp N(t)\rp$ given by
	\eqref{eq:eq_ch5_exp_stream_1} and $\E\lp X^{(1)}\rp=1/\lambda_1$.
\end{proof}

\subsubsection{Lower Bound on the Average Age of Stream $\mathcal{U}_1$}
We now compute a lower bound of the average age. 

Consider a fictitious system where if a stream $\mathcal{U}_1$ arrival finds the system in state $q'_1$, then the stream
$\mathcal{U}_2$ packet that is being served is discarded (and the stream $\mathcal{U}_1$ packet enters service immediately).
The instantaneous age process of this fictitious system is pointwise less than the instantaneous age of the true system,
consequently its average age lower bounds the true average age. Note that the fictitious system from the point of view of the
stream $\mathcal{U}_1$ is M/G/1, with service time distributed like $Y$ in \eqref{eq:eq_ch5_mgf_Y}.

\begin{lemma}
	\label{lemma:lemma_ch5_system_time_lb}
	Assume an M/G/1 queue with interarrival time $X^{(1)}$ exponentially distributed with rate $\lambda_1$ and service time
	$Y$ whose moment generating function is given by \eqref{eq:eq_ch5_mgf_Y}. The service time and the interarrival time are
	assumed to be independent. Then the distribution of the system time $T$ is 
	{\small\begin{equation}
		\label{eq:eq_ch5_dist_system_time_lb}
		f_T(t) = C_1 e^{-\alpha_1 t}(\mu_2-\alpha_1) - C_1 e^{-\alpha_2 t}(\mu_2-\alpha_2),\ t\geq0,
	\end{equation}}%
	where $\alpha_1,\alpha_2>0$ are the roots of the quadratic expression
	$$s^2-s(\mu_1+\mu_2+\lambda_2-\lambda_1)+\mu_1\mu_2-\lambda_1\mu_2-\lambda_1\lambda_2,$$
	$$C_1 = \frac{(1-\rho)\mu_1}{\alpha_2-\alpha_1},$$
	and $\rho = \lambda_1\E\lp Y\rp =
	\frac{\lambda_1(\mu_2+\lambda_2)}{\mu_1\mu_2}$.
\end{lemma}
\begin{proof}
	See Appendix~\ref{appendix:appendix_proof_lemma_ch5_system_time_lb}.
\end{proof}
From \cite{KaulYatesGruteser-2012Infocom}, we know that the average age of the M/G/1 queue with interarrival time $X^{(1)}$ and
service time $Y$ is 
\begin{align}
	\label{eq:eq_ch5_avg_age_lb}
	\Delta_{LB} &= \lambda_1\lp\frac{1}{2}\E\lp {X^{(1)}_j}^2\rp + \E\lp T_jX^{(1)}_j\rp\rp,
\end{align}
where for the $j^{th}$ packet we have $T_j = (T_{j-1}-X^{(1)}_j)^+ + Y_j$, $f(x) =(x)^+ = x\mathbbm{1}_{\{x\geq 0\}}$ and
$\mathbbm{1}_{\{.\}}$ is the
indicator function. So $\E\lp T_jX^{(1)}_j\rp$ becomes 
\begin{equation}
	\label{eq:eq_ch5_ETX}
	\E\lp T_jX^{(1)}_j\rp = \E\lp X^{(1)}_j(T_{j-1}-X^{(1)}_j)^+\rp + \E\lp Y_j\rp\E\lp X^{(1)}_j\rp,
\end{equation}
where the second term is due to the independence between $Y_j$ and $X^{(1)}_j$.

\begin{proposition}
	\label{prop:prop_ch5_ETXX}
\begin{align}
	\label{eq:eq_ch5_ETXX}
	&\E\lp X^{(1)}_j(T_{j-1}-X^{(1)}_j)^+\rp \nn
	&= \frac{\lambda_1\mu_2+2\lambda_1\lambda_2}{\mu_1^2(\mu_1\mu_2-\lambda_1(\mu_2+\lambda_2))}\nn 
	&+
	\frac{\lambda_2\lambda_1}{\mu_2}\lp
	\frac{(\mu_2+\mu_1+\lambda_2)^2-2\mu_1\mu_2}{\mu_1^2(\mu_2+\lambda_1)(\mu_1\mu_2-\lambda_1(\mu_2+\lambda_2))}
\right)\nn
	&+ \frac{\lambda_2\lambda_1}{\mu_2}\left(
\frac{2\mu_2\lambda_1(\mu_1+\lambda_2)+\lambda_2(\lambda_1^2+\mu_2)}{\mu_1^2(\mu_2+\lambda_1)^2(\mu_1\mu_2-\lambda_1(\mu_2+\lambda_2))}\rp.
\end{align}
\end{proposition}
\begin{proof}
	Given that $T_{j-1}$ and $X^{(1)}_j$ are independent then 
	\begin{align*}
		&\E\lp X^{(1)}_j(T_{j-1}-X^{(1)}_j)^+\rp\nn
		&= \int_0^\infty \int_x^\infty x(t-x)f_T(t)\lambda_1 e^{-\lambda_1 x}\mathrm{d}t\mathrm{d}x
	\end{align*}
	Replacing $f_T(t)$ by its value in \eqref{eq:eq_ch5_dist_system_time_lb} and using the fact that 
	\begin{align*}
	\alpha_1+\alpha_2 = \mu_1+\mu_2+\lambda_2-\lambda_1,\nn
	\alpha_1\alpha_2 = \mu_1\mu_2-\lambda_1\mu_2-\lambda_1\lambda_2,
	\end{align*}
	we get \eqref{eq:eq_ch5_ETXX} after some computations.
\end{proof}
\begin{thm}
	\label{thm:thm_ch5_lower_bound_avg_age}
			\begin{align}
				\label{eq:eq_ch5_lower_bound_avg_age}
			\Delta_{LB} &= \frac{1}{\lambda_1} + \frac{\mu_2+\lambda_2}{\mu_1\mu_2}+
\frac{\lambda_1^2\mu_2+2\lambda_1^2\lambda_2}{\mu_1^2(\mu_1\mu_2-\lambda_1(\mu_2+\lambda_2))}\nn
				    &+\frac{\lambda_2\lambda_1^2}{\mu_2}\lp\frac{(\mu_2+\mu_1+\lambda_2)^2-2\mu_1\mu_2}{\mu_1^2(\mu_2+\lambda_1)(\mu_1\mu_2-\lambda_1(\mu_2+\lambda_2))}\right) \nn
	&+\frac{\lambda_2\lambda_1^2}{\mu_2}\left(\frac{2\mu_2\lambda_1(\mu_1+\lambda_2)+\lambda_2(\lambda_1^2+\mu_2)}{\mu_1^2(\mu_2+\lambda_1)^2(\mu_1\mu_2-\lambda_1(\mu_2+\lambda_2))}\rp.
		\end{align}
This is also a lower bound on the true average age of stream $\mathcal{U}_1$ packets.
\end{thm}
\begin{proof}
Using \eqref{eq:eq_ch5_ETXX}, $\E\lp Y_j\rp=\E\lp Y\rp=\frac{\mu_2+\lambda_2}{\mu_1\mu_2}$ and $\E\lp X^{(1)}_j\rp=\E\lp
X^{(1)}\rp = \frac{1}{\lambda_1}$, we can find a closed-form expression for $\E\lp T_jX^{(1)}_j\rp$. Replacing this expression
in \eqref{eq:eq_ch5_avg_age_lb} and using the fact that $\E\lp {X^{(1)}_j}^2\rp= \frac{2}{\lambda_1^2}$, we obtain a closed-form
expression of the average age $\Delta_{LB}$ of an M/G/1 queue with interarrival time $X^{(1)}$ and service time $Y$. 
\end{proof}


\subsubsection{Average Age of Stream $\mathcal{U}_2$}
\label{subsec:subsec_ch5_age_stream_1}
By design, stream $\mathcal{U}_2$ is not interfered at all by
stream $\mathcal{U}_1$ hence behaves like a
traditional M/M/1/1 with preemption queue with generation rate $\lambda_2$ and service rate $\mu_2$. The average age of this
stream was computed in \cite{2012CISS-KaulYatesGruteser} to be 
\begin{equation}
	\label{eq:eq_ch5_stream_2_age}
	\Delta_{\mathcal{U}_2} = \frac{1}{\mu_2}+\frac{1}{\lambda_2}.
\end{equation}


%
%
%
%
%
%
%


\section{M/G/1/1 with Preemption for the Low-Priority Stream}
\label{sec:sec_ch5_preempt_low_priority}
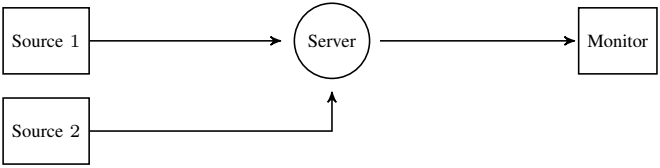
\begin{figure}
   \centering
   	\begin{tikzpicture}[>=stealth',shorten >=1pt,auto,node distance=3.8cm,semithick,font=\scriptsize]
	\tikzstyle{every state}=[rectangle,fill=white,draw=black,text=black]
	\node[state] (A)					{Source $1$};
	\node[state] (B) [node distance=1.2cm,below of=A]	{Source $2$};

	\node (C) [right of=A]	{\begin{tikzpicture}[>=stealth',shorten >=1pt,auto,semithick]
		\draw (0,0) circle (0.5cm) node {Server};\end{tikzpicture}};
	\node[state] (D) [right of=C]				{Monitor};
	
	\path[->] (A) edge node[] {} (C); 
	\draw[->] (B.east)-- ++(0,0)-| (C.south);
	\path[->] (C) edge node[] {} (D); 
		
	\end{tikzpicture}
	\caption{Diagram representing the model with preemption for the low priority stream.}
	\label{fig:fig_ch5_preempt_model}
\end{figure}
Fig.~\ref{fig:fig_ch5_preempt_model} presents an illustration of the model. In this model, we assume we have no memory, hence packets from stream $\mathcal{U}_1$ preempt each other. However, if an
arriving $\mathcal{U}_1$ packet finds the system busy serving a $\mathcal{U}_2$ packet, the server discards the stream
$\mathcal{U}_1$ packet because stream $\mathcal{U}_2$ packets are given higher priority. Furthermore, the server applies 
a preemption policy whenever a packet from $\mathcal{U}_2$ is generated. This means that if a newly generated packet from stream 
$\mathcal{U}_2$ finds the system busy (serving a packet from $\mathcal{U}_1$ or $\mathcal{U}_2$), the server preempts the
update currently in service and starts serving the new packet. Moreover, if the preempted packet belongs to $\mathcal{U}_1$ or
$\mathcal{U}_2$, this packet is discarded. 

These ideas are illustrated in part in Fig.~\ref{fig:fig_ch5_low_preempt_fig1}, which also shows the variation of the instantaneous age of stream
$\mathcal{U}_1$. In this plot, $t_j$ refers to the generation time of the $j^{th}$ packet, and $D_i$ corresponds to the
delivery time of the $i^{th}$ successfully received packet of stream
$\mathcal{U}_1$. As in this case not all the packets generated by source $\mathcal{U}_1$ are received, we distinguish
between \emph{generated} packets and \emph{successful} packets. Moreover, $t'_i$ and $D'_i$ are the start and end times of the $i^{th}$ period during which the system is busy
serving packets only from stream $\mathcal{U}_2$.  


\subsection{Ages of Streams\ $\mathcal{U}_1$ and\ $\mathcal{U}_2$}
\label{subsubsec:subsubsec_ch5_low_preempt_low_preempt_ages_stream_1_2}

\subsubsection{Preliminaries}
\label{subsubsec:subsubsec_ch5_low_preempt_preliminaries}
In this section also, unless stated otherwise, all random variables correspond to stream $\mathcal{U}_1$. We also follow the
convention where a random variable $U$ with no subscript corresponds to the steady-state version of $U_j$ that refers to the
random variable relative to the $j^{th}$ received packet from stream $\mathcal{U}_1$. To differentiate between streams, we
use superscripts, so that $U^{(i)}$ corresponds to the steady-state variable $U$ relative to stream $\mathcal{U}_i$ , $i=1,2$. 

In contrast to \Cref{sec:sec_ch5_fcfs_low_priority}, here we follow a slightly different notation: 
\begin{itemize}
	\item $X^{(i)}$ is the interarrival time between two consecutive generated updates from stream
$\mathcal{U}_i$, so $f_{X^{(i)}}(x) = \lambda_i e^{-\lambda_i x}$, $i=1,2$,
	\item $S^{(i)}$ is the
	service time random variable of stream $\mathcal{U}_i$ updates with p.d.f $f_{S^{(i)}}(t)$, $i=1,2$,
	\item $T_j$ is the system time, or the time spent
by the $j^{th}$ successfully received stream $\mathcal{U}_1$ update in the queue,
	\item $Y_j$ to be the interdeparture time between the $j^{th}$ and $j+1^{th}$ successfully received stream $\mathcal{U}_1$ updates. 
	\item $R(\tau)=\max\left\{n:D_{n}\leq\tau\right\}$ is the number of  successfully received updates from stream $1$ in
		the interval $[0,\tau]$. 
\end{itemize}
Given that in this model there is no waiting
in the queue, the system time of a received packet is equal to its service time. In our model, we assume the service time of the 
updates from the different streams to be independent of the interarrival time between consecutive packets (regardless if they belong to the same
stream).

Finally, two important quantities that we will use extensively are 
\begin{itemize}
\item $P_\lambda = \mathbb{E}\lp e^{-\lambda S^{(1)}}\rp= \int f_{S^{(1)}}(t) e^{-\lambda t} \mathrm{d}t,$
\item $L_{\lambda_2} = \mathbb{E}\lp e^{-\lambda_2 S^{(2)}}\rp= \int f_{S^{(2)}}(t) e^{-\lambda_2 t} \mathrm{d}t.$
\end{itemize} 
These are the Laplace transform of $f_{S^{(1)}}(t)$ and $f_{S^{(2)}}(t)$ evaluated at $\lambda=\lambda_1+\lambda_2$ and
$\lambda_2$, respectively.
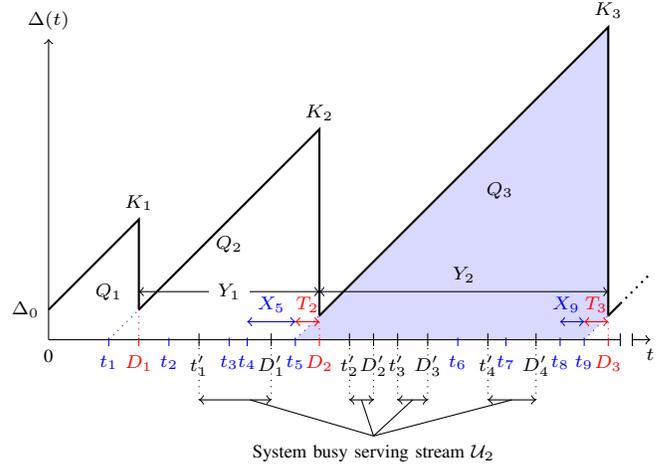
\begin{figure}[t]
	\centering
	\begin{tikzpicture}[scale=0.8,font=\scriptsize]
		\path[fill=blue!15] (4.1,0)--(9.3,5.2)--(9.3,0.4)--(8.9,0)--cycle;
		\draw (0,0) -- (9.5,0);
		\draw (9.5,3pt) -- (9.5,-3pt);
		\draw (9.7,3pt) -- (9.7,-3pt);
		\draw[->] (9.7,0) -- (10,0)node[anchor=north] {$t$};
		\draw	(0,0) node[anchor=north] {0};
		\draw[color=blue]	(1,1pt) -- (1,-3pt) node[anchor=north] {{$t_1$}};
		\draw[color=red]	(1.5,1pt) -- (1.5,-3pt) node[anchor=north] {$D_1$};
		
		\draw[color=blue]	(2,1pt) -- (2,-3pt) node[anchor=north] {$t_2$};
		\draw			(2.5,3pt) -- (2.5,-3pt) node[anchor=north] {$t'_1$};
		\draw[color=blue]	(3,1pt) -- (3,-3pt) node[anchor=north] {$t_3$};
		\draw[color=blue]	(3.3,1pt) -- (3.3,-3pt) node[anchor=north] {$t_4$};
		\draw			(3.7,3pt) -- (3.7,-3pt) node[anchor=north] {$D'_1$};
		\draw[color=blue]	(4.1,1pt) -- (4.1,-3pt) node[anchor=north] {$t_5$};
		\draw[color=red]	(4.5,1pt) -- (4.5,-3pt) node[anchor=north] {$D_2$};
		
		\draw			(5,3pt) -- (5,-3pt) node[anchor=north] {$t'_2$};
		\draw			(5.4,3pt) -- (5.4,-3pt) node[anchor=north] {$D'_2$};
		\draw			(5.8,3pt) -- (5.8,-3pt) node[anchor=north] {$t'_3$};
		\draw			(6.3,3pt) -- (6.3,-3pt) node[anchor=north] {$D'_3$};
		\draw[color=blue]	(6.8,1pt) -- (6.8,-3pt) node[anchor=north] {$t_6$};
		
		\draw			(7.3,3pt) -- (7.3,-3pt) node[anchor=north] {$t'_4$};
		\draw[color=blue]	(7.6,1pt) -- (7.6,-3pt) node[anchor=north] {$t_7$};
		\draw			(8.1,3pt) -- (8.1,-3pt) node[anchor=north] {$D'_4$};
		\draw[color=blue]	(8.5,1pt) -- (8.5,-3pt) node[anchor=north] {$t_8$};
		\draw[color=blue]	(8.9,1pt) -- (8.9,-3pt) node[anchor=north] {$t_9$};
		\draw[color=red]	(9.3,1pt) -- (9.3,-3pt) node[anchor=north] {$D_3$};
					
		\draw[->] (0,0) -- (0,5) node[anchor=south] {$\Delta(t)$};
				
		\draw[thick] (0,0.5) -- (1.5,2) -- (1.5,0.5);
		\draw[dotted, color=blue] (1,0) -- (1.5,0.5);
		\draw[dotted, color=red]  (1.5,0.5) -- (1.5,0);
		\draw (0,0.5) node[anchor=east] {$\Delta_0$}; 
		\draw (1.5,2) node[anchor=south] {$K_1$};
		\draw (1,0.5) node[anchor=south] {$Q_1$};
		
		\draw[thick] (1.5,0.5) -- (4.5,3.5) -- (4.5,0.4);
		\draw[dotted, color=blue] (4.1,0) -- (4.5,0.4);
		\draw[dotted, color=red]  (4.5,0.4) -- (4.5,0);
		\draw (4.5,3.5) node[anchor=south] {$K_2$};
		\draw (3,1.3) node[anchor=south] {$Q_2$};
			
		\draw[thick] (4.5,0.4) -- (9.3,5.2) -- (9.3,0.4);
		\draw[dotted, color=blue] (8.9,0) -- (9.3,0.4);
		\draw[dotted, color=red]  (9.3,0.4) -- (9.3,0);
		\draw (9.3,5.2) node[anchor=south] {$K_3$};
		\draw (7.5,2.2) node[anchor=south] {$Q_3$};
		
		\draw[thick] (9.3,0.4) -- (9.5,0.6);
		\draw[dotted, thick] (9.5,0.6)--(10,1.1);

		\draw[<->] (2.5,-1) -- (3.7,-1);
		\draw[<->] (5,-1) -- (5.4,-1);
		\draw[<->] (5.8,-1) -- (6.3,-1);
		\draw[<->] (7.3,-1) -- (8.1,-1);
		
		\draw[dotted] (2.5,-1) -- (2.5,0);
		\draw[dotted] (3.7,-1) -- (3.7,0);
		\draw[dotted] (5,-1) -- (5,0);
		\draw[dotted] (5.4,-1) -- (5.4,0);
		\draw[dotted] (5.8,-1) -- (5.8,0);
		\draw[dotted] (6.3,-1) -- (6.3,0);
		\draw[dotted] (7.3,-1) -- (7.3,0);
		\draw[dotted] (8.1,-1) -- (8.1,0);

		\draw (3.35,-1)--(5.4,-1.6);
		\draw (5.2,-1)--(5.4,-1.6);
		\draw (6.05,-1)--(5.4,-1.6);
		\draw (7.7,-1)--(5.4,-1.6) node[anchor=north] {System busy serving stream $\mathcal{U}_2$};
		
		\draw[<->, color=red] (8.9,0.3) -- (9.3,0.3);
		\draw[color=red] (9.1,0.3) node[anchor=south, fill=none] {$T_3$};
		\draw[<->, color=red] (4.1,0.3) -- (4.5,0.3);
		\draw[color=red] (4.3,0.3) node[anchor=south, fill=none] {$T_2$};
		\draw[<->] (4.5,0.8) -- (9.3,0.8);
		\draw (6.9,0.8) node[anchor=south, fill=none] {$Y_2$};

		\draw[<->] (1.5,0.8) -- (4.5,0.8);
		\draw (3,0.8) node[anchor=center, fill=white] {$Y_1$};
		\draw[<->, color=blue] (3.3,0.3) -- (4.1,0.3);
		\draw[color=blue] (3.7,0.3) node[anchor=south, fill=none] {$X_5$};
		\draw[<->, color=blue] (8.5,0.3) -- (8.9,0.3);
		\draw[color=blue] (8.6,0.3) node[anchor=south, fill=none] {$X_9$};

	\end{tikzpicture}
	\caption{Variation of the instantaneous age of stream $\mathcal{U}_1$.}
	\label{fig:fig_ch5_low_preempt_fig1}
\end{figure}
\subsubsection{Average Age and Average Peak age of Stream $\mathcal{U}_1$}
\label{subsubsec:subsubsec_ch5_low_preempt_age_stream_2}
\begin{lemma}
	\label{lemma:lemma_ch5_low_preempt_mg11_sys_time}
	For the priority preemption system described above, the moment generating function of the system time $T$
	corresponding to stream $\mathcal{U}_1$ is given by
	\begin{equation}
		\label{eq:eq_ch5_low_preempt_mg11_T}
		\phi_{T}(s) = \frac{P_{\lambda-s}}{P_\lambda}.
	\end{equation}
\end{lemma}
\begin{proof}
	All variables in this proof corresponds to stream $\mathcal{U}_1$. The system time $T_j$ of
	the $j^{th}$ successfully received packet corresponds to the service time of the $j^{th}$ received packet given that
	service was completed before any new arrival (because any new packet from any stream will preempt the current update
	being served). Therefore, in steady state, $\Prob\lp T>t\rp = \Prob\lp S^{(1)}>t|S^{(1)}<\min\lp X^{(1)}, X^{(2)}\rp\rp$. Hence, for
	$L = \min\lp X^{(1)}, X^{(2)}\rp$,
	\begin{align*}
		f_T(t) &= \lim_{\epsilon\to 0}\frac{\Prob\lp T\in [t,t+\epsilon] \rp}{\epsilon}\\
		       &= \lim_{\epsilon\to 0}\frac{\Prob\lp S^{(1)}\in [t,t+\epsilon]|S^{(1)}<L\rp}{\epsilon}\\
		       &= \lim\limits_{\epsilon\to 0}\resizebox{0.35\textwidth}{!}{$\frac{\Prob\lp S^{(1)}\in [t,t+\epsilon]\rp \Prob\lp S^{(1)}<L| S^{(1)}\in
	       [t,t+\epsilon]\rp}{\epsilon\Prob\lp S^{(1)}<L\rp}$}\\
		       &= \frac{f_{S^{(1)}}(t)\Prob\lp L>t\rp}{\Prob\lp S^{(1)}<L\rp} = \frac{f_{S^{(1)}}(t)e^{-\lambda t}}{\Prob\lp S^{(1)}<L\rp},
	\end{align*}
	where the last equality is due to the fact that $L$ is exponentially distributed with rate $\lambda=\lambda_1+\lambda_2$. Thus,
	\begin{align*}
		\phi_T(s) = \E\lp e^{sT} \rp &= \int_0^\infty \frac{f_{S^{(1)}}(t)}{\Prob\lp
		S^{(1)}<L\rp}e^{-(\lambda-s)t}\mathrm{d}t\nn 
		&= \frac{P_{\lambda-s}}{\Prob\lp S^{(1)}<L\rp}.
	\end{align*}
	Finally,
	\begin{align*}
		\Prob\lp S^{(1)}<L\rp &= \int_0^\infty f_{S^{(1)}}(t)\Prob\lp L>t\rp\mathrm{d}t\nn
				      &= \int_0^\infty f_{S^{(1)}}(t)e^{-\lambda t}\mathrm{d}t= P_\lambda.
	\end{align*}
\end{proof}

\begin{lemma}
	\label{lemma:lemma_ch5_low_preempt_mg11_Y}
	The moment generating function of the interdeparture time of stream $\mathcal{U}_1$, $Y$, is 
	\begin{equation}
		\label{eq:eq_ch5_low_preempt_mg11_Y}
		\phi_{Y}(s) = \frac{\lambda_1 P_{\lambda-s}\lp \lambda_2 L_{\lambda_2-s}-s\rp}{\lambda_1 P_{\lambda-s}\lp
\lambda_2 L_{\lambda_2-s} -s\rp-s(\lambda_2-s)}. 
	\end{equation}
\end{lemma}

\begin{proof}
	We use again the detour flow graph method. The detailed proof is presented in
	Section~\ref{appendix:appendix_proof_lemma_ch5_low_preempt_mg11_Y}.	
\end{proof}
Before presenting the main theorem of this section, we need the following lemma that proves that the studied system is ergodic.
\begin{lemma}
	\label{lemma:lemma_ch3_T_Y_independ}
	Consider stream $\str$. For any $j\geq1$, the random variables $T_j$ and $Y_j$ relative to the $j^{th}$ successful packet are independent.
	Moreover the process $(Y_j)_{j\geq1}$ is i.i.d, with its distribution given by
\Cref{lemma:lemma_ch5_low_preempt_mg11_Y}, and the process $R(\tau)=\sup\{n\in\mathbb{N};D_n\leq\tau\}$ is a renewal process. 
\end{lemma}
\begin{proof}
Let $L_j=\min\lp X^{(1)}_j,X^{(2)}\rp$. Since the interarrival times for both streams are exponential and independent, $L_j$ is also
exponential with rate $\lambda=\lambda_1+\lambda_2$. Except $L_j$, all other variables are relative to stream $\str$. The
$j^{th}$ successful packet leaves the queue empty hence
$Y_j=\hat{X}_j+Z_j$. $\hat{X}_j=L_j-T_j$ is the remaining time between the departure of the stream-$\str$ $j^{th}$
successful packet, and the generation time of the next packet to be transmitted (it can belong to stream $\str$ or stream
$\strr$). $Z_j$ is the time for a new stream-$\str$ packet to be successfully delivered. $Z_j$ does not overlap with $T_j$ and
thus is independent from it. As for $\hat{X}_j$, we also obtain that it is independent of $T_j$. Intuitively, since $L_j$ is
exponentially distributed with rate $\lambda$ and thus memoryless, then the distribution of the remainder $\hat{X}_j$ is independent of the value
of $T_j$. Indeed, $\hat{X}_j$ is also exponentially distributed with rate $\lambda$. A more formal proof can be found in
Appendix~\ref{appendix:appendix_proof_lemma_ch3_T_Y_independ}. 

Furthermore, since $Y_{j-1}=\hat{X}_{j-1}+Z_{j-1}$, $\hat{X}_j$ is independent from $T_j$ and the interarrival process is i.i.d
and independent from the i.i.d service process, then $\hat{X}_j$  and $Z_j$ are independent of $Y_{j-1}$. This implies that for
any $j\geq1$, $Y_{j-1}$ and $Y_j$ are independent. Moreover, it is clear that the $Z_j$'s have the same distribution. Since the
$\hat{X}_j$'s are exponential with rate $\lambda$ then the $(Y_j)_{j\geq1}$ is an i.i.d
process. Given that $Y_j$ is the interval of time between the receptions of two consecutive successful stream-$\str$ packets, then the
number of successfully received packets in the interval $[0,\tau]$, $R(\tau)$, is a renewal process.
\end{proof}
Now we can state the main theorem of this section.
\begin{thm}
	\label{thm:thm_ch5_low_preempt_mg11_age}
	Assume an M/G/1/1 queue with preemption and a sender consisting of two sources generating packets according to two
	independent Poisson processes with rates $\lambda_i$, $i=1,2$, such that $\lambda=\lambda_1+\lambda_2$. Moreover, packets
	belonging to stream $i$ are served according to $S^{(i)}$. If stream $\strr$ is given higher priority over stream
	$\str$, then
	\begin{enumerate}
		\item the average age of stream $\str$ is given by
			\begin{equation}
				\label{eq:eq_ch5_low_preempt_avg_age_mg11}
				\Delta_1 = \frac{1}{\lambda_1 P_\lambda
L_{\lambda_2}}+\frac{1-L_{\lambda_2}-\lambda_2\mathbb{E}\lp \SsS e^{-\lambda_2\SsS}\rp}{\lambda_2 L_{\lambda_2}}
			\end{equation}
		\item and the average peak age of stream $\str$ is given by
			\begin{equation}
				\label{eq:eq_ch5_low_preempt_avg_peak_age_mg11}
				\Delta_{peak,1} = \frac{1}{\lambda_1 P_\lambda L_{\lambda_2}}+\frac{\E\left(\Ss e^{-\lambda
\Ss}\right)}{P_\lambda}.
			\end{equation}
	\end{enumerate}
\end{thm}
\begin{proof}
	
	By \Cref{lemma:lemma_ch3_T_Y_independ}, $R(\tau)$ forms a renewal process. By \cite{ross},
	$\lim_{\tau\to\infty}\frac{R(\tau)-1}{\tau}=\frac{1}{\E(Y)}$, where $Y$ is the steady-state interdeparture random
	variable. Introducing the quantity $C_j=\int_{D_j}^{D_{j+1}}\Delta(t)\mathrm{d}t$ to be the reward function over
	the renewal period $Y_j$, we obtain using renewal reward theory \cite{Gallager1996,ross} that
	\[\Delta_1 = \lim_{\tau\to\infty} \frac{1}{\tau}\int_0^\tau \Delta(t)\mathrm{d}t =
	\frac{\E(C_j)}{\E(Y_j)}=\frac{\E(Q_j)}{\E(Y_j)}=\frac{\E(Q)}{\E(Y)}<\infty,\]
	where $Q$ is the steady-state counterpart of $Q_j$, and the last equality stems from the fact that the average age for
	Stream $1$ can also be also expressed as the sum
	of the geometric areas $Q_j$ under the instantaneous age curve of Fig.~\ref{fig:fig_ch5_low_preempt_fig1}. 
	
	It was shown in \cite{KaulYatesGruteser2012q,CostaCodreanuEphremides2014ISIT,NajmYatesSoljanin-ISIT2017} that, using
	Fig.~\ref{fig:fig_ch5_low_preempt_fig1}, 
	$$ \E\left(Q\right) = \frac{1}{2} \E\lp Y^2\rp +\E\lp TY\rp.$$
	Since, by \Cref{lemma:lemma_ch3_T_Y_independ}, the variables $T_j$ and $Y_j$ are independent for any $j\geq1$, then
	$$ \E\left(Q\right)= \frac{1}{2} \E\lp Y^2\rp +\E\lp T\rp \E\lp Y\rp.$$ 
	Therefore,
	\begin{equation}
		\label{eq:eq_ch5_low_preempt_steady_state_mg11_age}
		\Delta_1 = \E\lp T\rp + \frac{\E\lp Y^2\rp}{2\E\lp Y\rp}
	\end{equation}
	Moreover, from Fig.~\ref{fig:fig_ch5_low_preempt_fig1} we see that the peak age at the instant before receiving the $j^{th}$ packet is given
	by $$K_j = T_{j-1}+Y_{j-1}.$$ Hence, at steady state we get 
	\begin{equation}
		\label{eq:eq_ch5_low_preempt_peak_age_1}
		\Delta_{peak,1}=\E\lp K\rp=\E\lp T\rp+\E\lp Y\rp.
	\end{equation}
	
	Using Lemma~\ref{lemma:lemma_ch5_low_preempt_mg11_sys_time}, we obtain $\E\lp T\rp= P_\lambda^{-1}\E\lp \Ss e^{-\lambda \Ss}\rp$. Using
	Lemma~\ref{lemma:lemma_ch5_low_preempt_mg11_Y}, we get that $\E\lp Y\rp= \lp \lambda_1 P_\lambda L_{\lambda_2}\rp^{-1}$ and 
	\begin{align*}
	\frac{\E\lp Y^2\rp}{2\E\lp Y\rp} &=-\frac{1}{\lambda_2}-\frac{\E\lp \Ss e^{-\lambda \Ss}\rp}{P_\lambda}-\frac{\E\lp
\SsS e^{-\lambda_2\SsS}\rp}{L_{\lambda_2}}\nn
					 &+\frac{1}{\lambda_1 P_\lambda	L_{\lambda_2}} +\frac{1}{\lambda_2 L_{\lambda_2}}.
	\end{align*}
	 Using these expressions in \eqref{eq:eq_ch5_low_preempt_steady_state_mg11_age} and
	 \eqref{eq:eq_ch5_low_preempt_peak_age_1}, we achieve our result for stream $\str$. 
\end{proof}
\subsubsection{Average Age of Stream $\mathcal{U}_2$}
\label{subsubsec:subsubsec_ch5_low_preempt_age_stream_1}
By design, stream $\mathcal{U}_2$ is not at all interfered by
stream $\mathcal{U}_1$ hence behaves like a
traditional M/G/1/1 with preemption queue with generation rate $\lambda_2$ and service time $\SsS$. The average age of this
stream was computed in \cite{NajmYatesSoljaninMG11,NajmTelatar2018} to be 
\begin{equation}
	\label{eq:eq_ch5_low_preempt_stream_2_age}
	\Delta_{_2} = \frac{1}{\lambda_2 L_{\lambda_2}}.
\end{equation}



\section{Discussion on the Age of the Low Priority Stream}
\label{sec:sec_discussion}
\begin{figure}[t]
	\centering
	\subfloat[$\mu_1=10$, $\mu_2=5$, $\lambda_1=2$ and $\lambda_2<\frac{\mu_1\mu_2}{\lambda_1}-\mu_2=20$. \label{fig:fig_ch5_bounds}]{\includegraphics[scale=0.4,trim={0 0.4cm 0 1.5cm},clip]{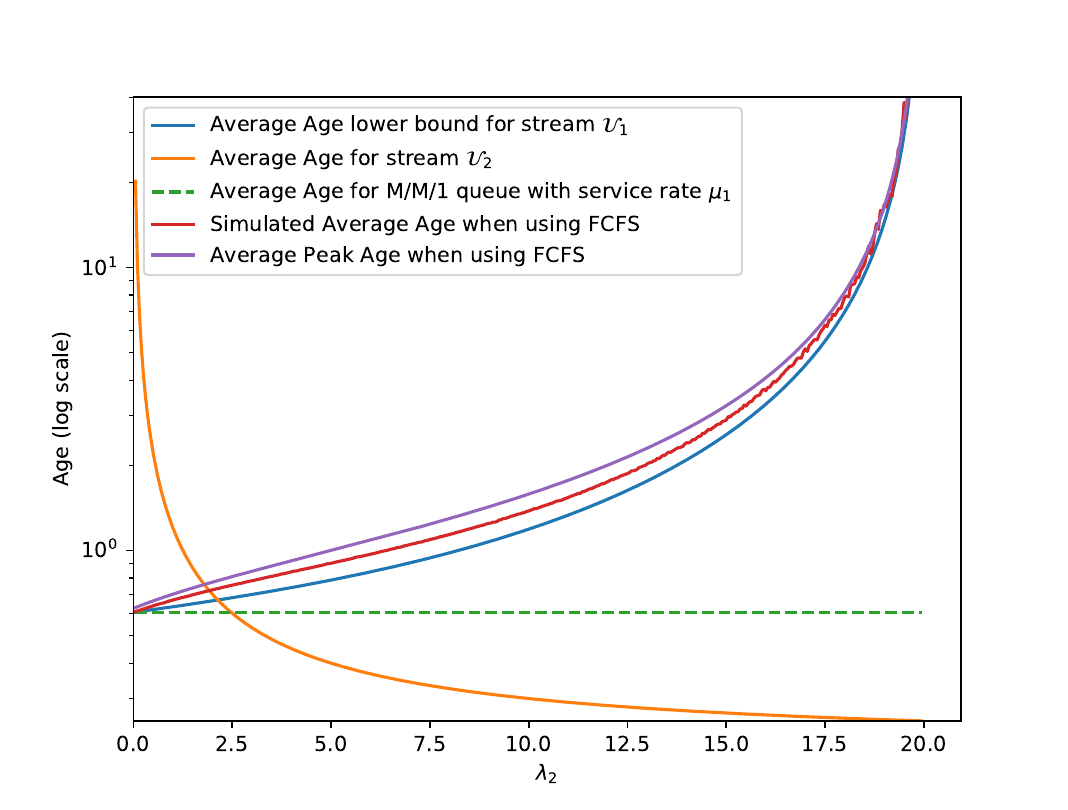}
}\\
	\subfloat[$\mu_1=6$, $\mu_2=18$, $\lambda_1=4$ and $\lambda_2<9$. \label{fig:fig_ch5_bounds_2}]{\includegraphics[scale=0.32,trim={0 0.5cm 0 1.5cm},clip]{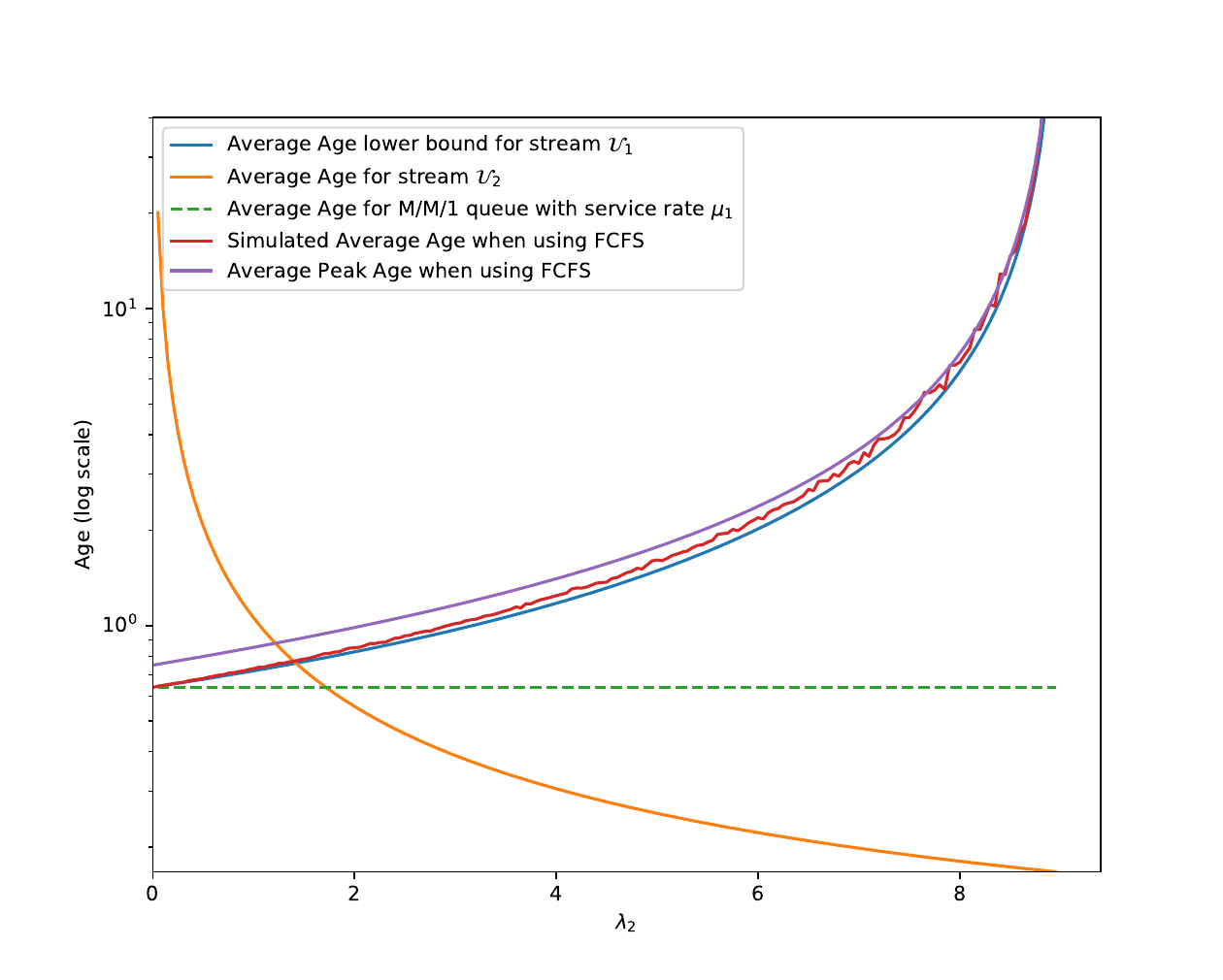}
}
	\caption{Plot of the  average age for stream $\mathcal{U}_2$ and average peak age and lower bound on the average age
	for stream $\mathcal{U}_1$.}
\end{figure}
Having analyzed the FCFS and the M/G/1/1/ with preemption transmission schemes for the low priority streams, we can now compare
their performances.
\subsection{Relative to the FCFS policy}
\label{sec:sec_ch5_numerical_results}

Fig.~\ref{fig:fig_ch5_bounds} shows the simulated average age, the average peak-age ($\Delta_{peak,1}$) and the lower bound on
the average age ($\Delta_{LB}$), as computed in the previous section
for stream $\mathcal{U}_1$, and the average age ($\Delta_{\mathcal{U}_2}$) of stream $\mathcal{U}_2$. In this plot, we fix $\mu_1=10$,
$\mu_2=5$, $\lambda_1=2$ and vary $\lambda_2$. As we can see, for stream $\mathcal{U}_1$ the average age,
the lower bound, and the average peak-age grow without bounds when $\lambda_2$ gets close to $\frac{\mu_1\mu_2}{\lambda_1}-\mu_2$. This
observation is in line with our result in Theorem~\ref{thm:thm_ch5_stat_dist} and the stability condition
\eqref{eq:eq_ch5_stab_condition}. In this simulation, we also notice
that the average peak age and the lower bound appear to be good bounds on the average age, especially for small $\lambda_2$ and
for values of $\lambda_2\sim 0$ close to the limit $\frac{\mu_1\mu_2}{\lambda_1}-\mu_2$.

It is easy to see via a coupling argument that if we increase $\lambda_2$, the age process $\Delta_{\mathcal{U}_1}(t)$ of the $\mathcal{U}_1$ stream will stochastically increase.  We see from the plots
that the lower bound on $\Delta_{\mathcal{U}_1}$ and that its average peak-age exhibit the same behavior. However, the
average age of stream $\mathcal{U}_2$ is decreasing in $\lambda_2$ (from \eqref{eq:eq_ch5_stream_2_age}). Consequently, minimizing $\Delta_{\mathcal{U}_2}$ and
minimizing $\Delta_{\mathcal{U}_1}$ are conflicting goals. 

We have seen that the average age of stream $\mathcal{U}_2$ is not affected by the presence of the other stream. However,
Fig.~\ref{fig:fig_ch5_bounds} shows the effect of stream $\mathcal{U}_2$ on the average age of stream $\mathcal{U}_1$
($\Delta_1$). For this, we
plot the average age ($\Delta_{ref}$) of an M/M/1 queue with generation rate $\lambda_1=2$ and service rate $\mu_1=10$ (given in
\cite{KaulYatesGruteser-2012Infocom}). We observe an expected behavior: for very low values of $\lambda_2$, the two average
ages and the lower bound $\Delta_{LB}$ are
close (they are all equal at $\lambda_2=0$). However, as $\lambda_2$ increases the presence of stream $\mathcal{U}_2$ quickly leads
to an increase in $\Delta_1$. In fact, for $\lambda_2=5$, $\Delta_1$ is already $50\%$ higher than $\Delta_{ref}$. This shows
that the presence of the priority stream $\mathcal{U}_2$ takes a heavy toll on the stream $\mathcal{U}_1$ age.
Another observation is that the average age curve of stream $\mathcal{U}_2$ crosses 
the average age of stream $\mathcal{U}_1$ at a value of $\lambda_2$, denoted $\lambda_2^*=1.9$.
This means that for $\lambda_2\leq\lambda_2^*$, stream $\mathcal{U}_2$ has an average age higher than stream $\mathcal{U}_1$. 
These observations show that not all values of $\lambda_2$ are suitable for our system. A small $\lambda_2$  will not ensure
for stream $\mathcal{U}_2$ the priority it needs, whereas a large $\lambda_2$ will make the average age of stream $\mathcal{U}_1$
large and the system unstable. 

Fig.~\ref{fig:fig_ch5_bounds_2} plots the same quantities as Fig.~\ref{fig:fig_ch5_bounds} but under different settings: in
this case, $\mu_1=6$, $\mu_2=18$, $\lambda_1=4$ and $\lambda_2<9$. In this particular scenario, we notice that the lower bound
is a tight bound on the simulated average age for all values of $\lambda_2$, and it is tighter than the average peak age.

\subsection{Relative to the M/G/1/1 with preemption policy}
\label{subsec:subsec_discussion_m/g/1/1_preemption}
A close observation of Equations \eqref{eq:eq_ch5_low_preempt_avg_age_mg11} and \eqref{eq:eq_ch5_low_preempt_avg_peak_age_mg11}
leads to the following remarks:
\begin{itemize}
\item If the service time for stream $\strr$ is $0$, $\Delta_1 = \frac{1}{\lambda_1 P_\lambda} \geq \frac{1}{\lambda_1
P_{\lambda_1}}$, where $\Delta = \frac{1}{\lambda_1 P_{\lambda_1}}$ is the value of the average of stream $\str$ if stream
$\strr$ is not present. This result is due to the fact that whenever a stream $\strr$ packet is generated, it immediately
preempts the stream $\str$ packet being served hence increases the instantaneous age of the latter stream. 
\item By using L'Hopital's rule, we can show that $$\lim_{\lambda_2\to 0} \Delta_1 = \Delta =\frac{1}{\lambda_1
P_{\lambda_1}}$$ as it is expected. The average peak-age also converges to its value when no stream $\strr$ exists.
\item \emph{Special case}: assume $\Ss\sim \text{Exp}(\mu_1)$ and $\SsS\sim \text{Exp}(\mu_2)$. Then 
	\begin{equation}
		\label{eq:eq_ch5_low_preempt_avg_age}
		\resizebox{0.45\textwidth}{!}{$\Delta_1 =
			\frac{(\mu_1+\lambda_1)}{\lambda_1\mu_1}\lp\frac{\mu_2+\lambda_2}{\mu_2}\rp+\frac{\lambda_2}{\mu_2}\lp\frac{\mu_2+\lambda_2}{\lambda_1\mu_1}+\frac{1}{\mu_2+\lambda_2}\rp$}
\end{equation}
and
\begin{equation}
	\label{eq:eq_ch5_low_preempt_avg_peak_age}
\Delta_{peak,1} = \frac{1}{\mu_1+\lambda_1+\lambda_2}+\frac{(\mu_1+\lambda_1+\lambda_2)(\mu_2+\lambda_2)}{\lambda_1\mu_1\mu_2}.
\end{equation}
Equation \eqref{eq:eq_ch5_low_preempt_avg_age} coincides exactly to the result obtained by Kaul et al. in
\cite{KaulYates-priorityISIT18} for the stream with lowest priority and when we have two sources. 

Denoting $\Delta_{Norm} = \frac{\mu_1+\lambda_1}{\mu_1\lambda_1}$ to be the average age of stream $\str$ when stream $\strr$
does not exist (see \cite{KaulYatesGruteser2012q}), we can compute the additional age the presence of stream $\strr$ costs to stream $\str$:
\begin{align*}
\Delta_{diff}&= \Delta_1 - \Delta_{Norm}\nn
&=\resizebox{0.37\textwidth}{!}{$\frac{\lambda_2}{\lambda_1\mu_1}+\frac{\lambda_2}{\lambda_1\mu_2}+\frac{\lambda_2}{\mu_1\mu_2}+\frac{\lambda_2^2}{\lambda_1\mu_1\mu_2}+\frac{\lambda_2}{\mu_2(\mu_2+\lambda_2)}.$}
\end{align*}
By letting $\mu_2\to\infty$ we obtain $\Delta_{diff}\to \frac{\lambda_2}{\lambda_1\mu_1}>0$, and by taking $\lambda_2=0$ we
obtain $\Delta_{diff}=0$ as predicted by the previous two remarks.
\end{itemize} 

\subsection{Comparing the two policies}
\label{subsec:subsec_discussion_two_policies}
\begin{figure}[t]
	\centering
	\subfloat[$\mu_1=10$, $\mu_2=5$, $\lambda_1=2$.\label{fig:fig_ch5_comparison_avg_peak_age_FCFS_vs_MG11}]{\includegraphics[scale=0.3]{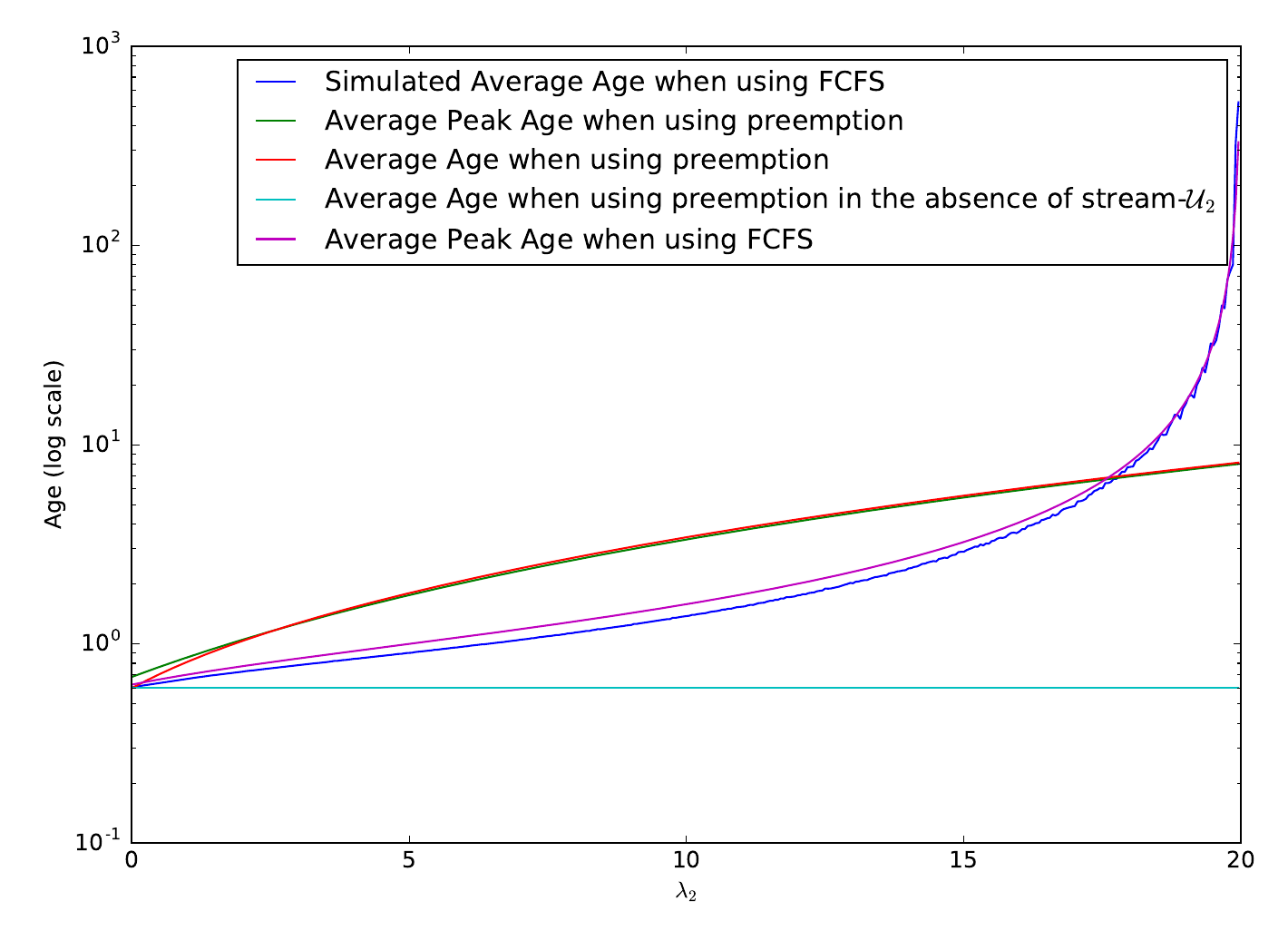}}\\
	\subfloat[$\mu_1=6$, $\mu_2=5$,	$\lambda_1=2$.\label{fig:fig_ch5_comparison_avg_peak_age_FCFS_vs_MG11_mu1_6}]{\includegraphics[scale=0.3]{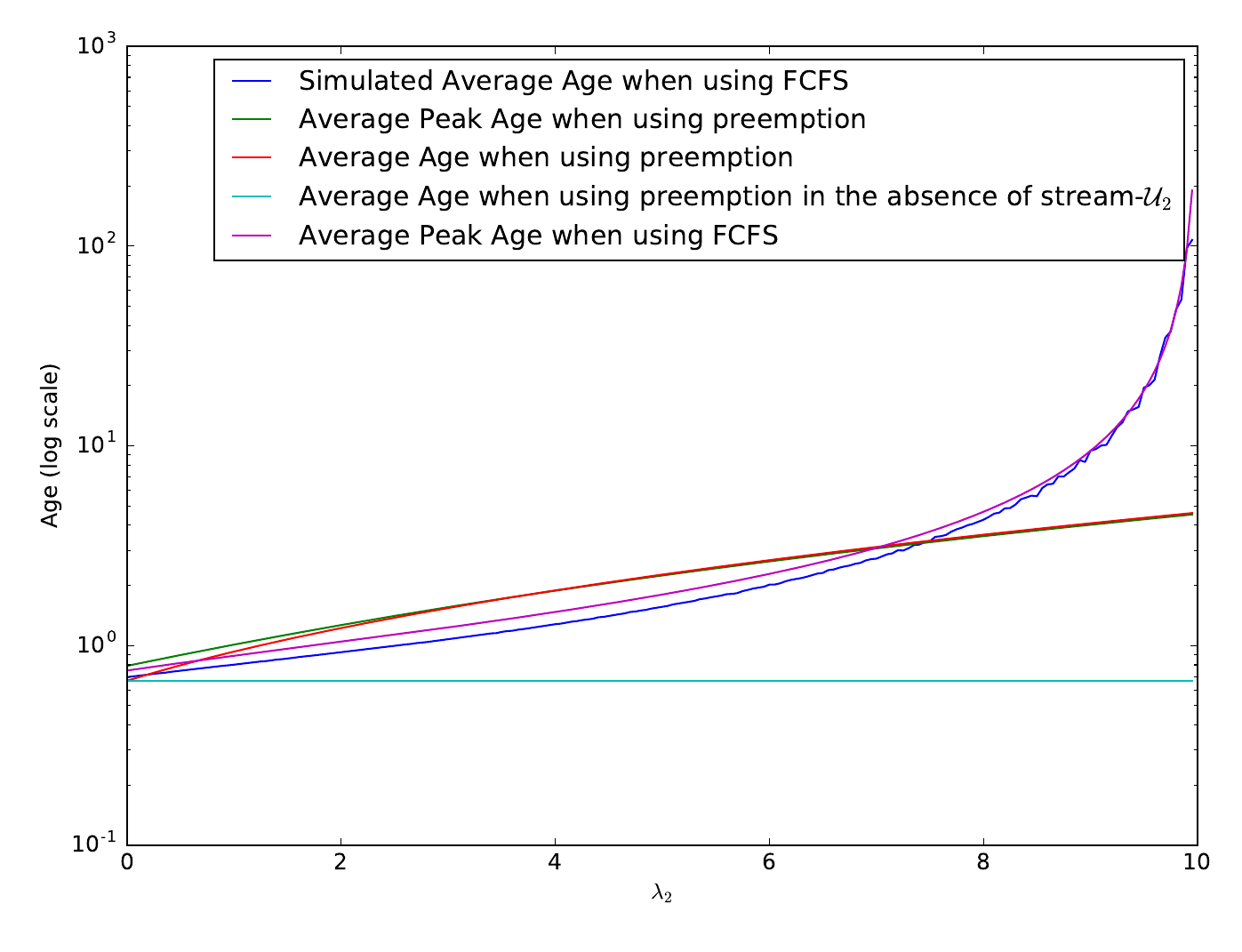}}\\
	\subfloat[$\mu_1=10$, $\mu_2=5$, $\lambda_1=4$.\label{fig:fig_ch5_comparison_avg_peak_age_FCFS_vs_MG11_l1_4}]{\includegraphics[scale=0.31]{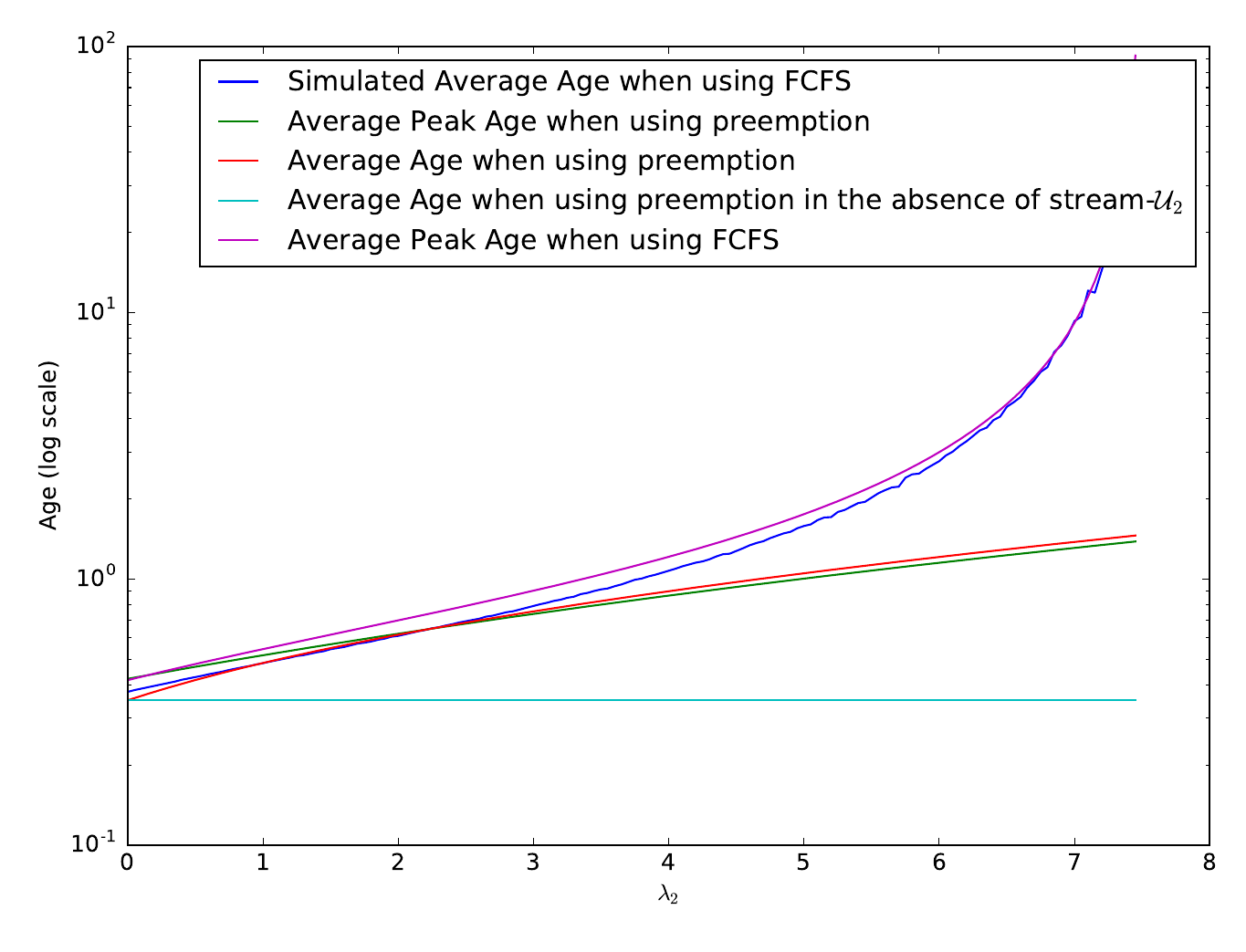}}
	\caption{Comparison between the average peak ages of the low priority source $\mathcal{U}_1$ when using the FCFS and the preemption
	schemes and assuming exponential service times.}
	\label{fig:fig_ch5_comparison_avg_peak_age_FCFS_vs_MG11_all}
\end{figure}
Using \eqref{eq:eq_ch5_avg_peak_age_1}, \eqref{eq:eq_ch5_low_preempt_avg_age} and
	\eqref{eq:eq_ch5_low_preempt_avg_peak_age}, we compare the performance
	of the preemption policy on stream $\str$ with that of the FCFS scheme from an age point of view when the service times
	corresponding to both sources are exponential. Fig.~\ref{fig:fig_ch5_comparison_avg_peak_age_FCFS_vs_MG11} plots the
	average ages and average peak-ages relative to stream $\str$ for the preemption, as well as for the FCFS schemes. In both
	cases, we assume stream $\str$ packets are generated according to a Poisson process of rate $\lambda_1= 2$ and served
	according to an exponential service time with rate $\mu_1=10$. As for stream $\strr$ updates, they are generated
	according to a Poisson process with rate $\lambda_2$ and served according to an exponential service time with rate
	$\mu_2=5$. We observe from Fig.~\ref{fig:fig_ch5_comparison_avg_peak_age_FCFS_vs_MG11} that the preemption scheme
	performs worse than the FCFS except when $\lambda_2$ is
	close to the FCFS stability condition. This observation comes as a surprise because we would think that the constraint
	of delivering all generated packets imposed by a FCFS system would pull the age up, compared to the more flexible
	preemptive scheme. However, we can explain this result in the following way: When using the preemptive scheme and
	not storing any updates, the system incurs a substantial idle time (from the source $\str$ point of view) during which
	it waits for a new stream-$\str$
	update to be generated. In fact, this is a direct consequence of the first remark in
	Section~\ref{subsec:subsec_discussion_m/g/1/1_preemption}. Moreover, Bedewy et al. in
	\cite{BedewySunShroff17} show that for a single source and exponential service time, the optimal policy to adopt is
	the preemptive scheme. Fig.~\ref{fig:fig_ch5_comparison_avg_peak_age_FCFS_vs_MG11} proves that the introduction of an
	additional source with higher priority has a significant impact on the performances of the different transmission schemes. For
	instance, the preemption scheme is not optimal anymore even for exponential service times. However, we can notice that
	for $\lambda_2$ very close to 0, the M/G/1/1 scheme has a lower average age than the FCFS scheme, which is to be
	expected.
	
	Fig.~\ref{fig:fig_ch5_comparison_avg_peak_age_FCFS_vs_MG11_mu1_6} and
	Fig.~\ref{fig:fig_ch5_comparison_avg_peak_age_FCFS_vs_MG11_l1_4} also plot the average age and average peak-age
	relative to stream $\str$ for both schemes (preemption and FCFS). Compared to
	Fig.~\ref{fig:fig_ch5_comparison_avg_peak_age_FCFS_vs_MG11}, in
	Fig.~\ref{fig:fig_ch5_comparison_avg_peak_age_FCFS_vs_MG11_mu1_6} we keep $\lambda_1 = 2$ and $\mu_2 = 5$ but decrease
	the service rate of stream $\str$ to $\mu_1 = 6$. Whereas for
	Fig.~\ref{fig:fig_ch5_comparison_avg_peak_age_FCFS_vs_MG11_l1_4}, we keep $\mu_1=10$ and $\mu_2=5$ but increase the
	generation rate of stream-$\str$ packets to $\lambda_1=4$. In
	Fig.~\ref{fig:fig_ch5_comparison_avg_peak_age_FCFS_vs_MG11_mu1_6}, we notice that by decreasing the service rate of
	stream $\str$, the performances of the FCFS scheme
	and the M/G/1/1 scheme get closer compared to Fig.~\ref{fig:fig_ch5_comparison_avg_peak_age_FCFS_vs_MG11}. We can
	explain this behavior in the following way: the performances exhibited by the two transmission schemes in
	Fig.~\ref{fig:fig_ch5_comparison_avg_peak_age_FCFS_vs_MG11_mu1_6} are worse than their respective counterparts in
	Fig.~\ref{fig:fig_ch5_comparison_avg_peak_age_FCFS_vs_MG11} because every transmitted packet needs more time on average
	to be serviced. However, the FCFS scheme is more affected by this degradation in service than the M/G/1/1 scheme due to
	the compound effect of a slower service time on the waiting time of the packets in the queue. This means that the
	packets waiting in the queue will sustain a higher waiting time on average which will affect the age. 

	In Fig.~\ref{fig:fig_ch5_comparison_avg_peak_age_FCFS_vs_MG11_l1_4}, we notice that by increasing the generation rate of
	stream $\str$, the performances of the M/G/1/1 scheme becomes better than that of the FCFS scheme for $\lambda_2\leq 1$
	and $\lambda_2\geq 2.5$, while the two performances are very close in the interval $1\leq\lambda_2\leq 2.5$. This shows
	that $\lambda_1$ increases, preemption performs better than FCFS because of two simultaneous effects: 
	\begin{enumerate}
		\item The idle time incurred by the M/G/1/1/ system waiting for a new packet is reduced. Hence the impact of
			the presence of stream $\strr$, as explained in the first remark of
			Section~\ref{subsec:subsec_discussion_m/g/1/1_preemption}, is decreased.
		\item As $\lambda_1$ increases, the queue for the FCFS system becomes more congested. This leads to an increase
			of the average waiting time sustained by the packets. 
	\end{enumerate}
	These two effects explain why we notice a paradigm shift and why in this case it is better to adopt an M/G/1/1 scheme.

To sum up, Fig.~\ref{fig:fig_ch5_comparison_avg_peak_age_FCFS_vs_MG11_all} shows that, for Poisson generation process and
exponential service time, the M/G/1/1 scheme is not optimal anymore and that the FCFS scheme might perform better depending on
the values of $\lambda_1$, $\lambda_2$, $\mu_1$ and $\mu_2$.

\section{Conclusion}

In this paper, we have studied the effect of implementing content-dependent policies on the average age of the packets. We
have considered
a sender that generates two independent Poisson streams with one stream having higher priority than the
other stream. The \lq\lq high priority\rq\rq\ stream is sent using a preemption policy, whereas at first the \lq\lq regular\rq\rq\ stream
is transmitted using a FCFS policy and then it is transmitted using preemption. We
derived the stability condition for the former system, as well as
closed-form expressions for the average peak-age and a lower bound on the average age of the \lq\lq regular\rq\rq\ stream. For
the latter system we have given exact expressions for the average age and average peak-age of the \lq\lq regular\rq\rq\
stream and we have shown through simulations that, even if the service times relative to both streams are exponential, preemption is
not the optimal strategy to adopt for the \lq\lq regular\rq\rq\ stream. In fact, for some fixed service rates and \lq\lq
regular\rq\rq\ stream generation rate, the FCFS strategy performs better for a large interval of \lq\lq high priority\rq\rq-
stream generation rate. 


\section*{Acknowledgements}

This research was supported in part by grant No.  200021\_166106/1 of the Swiss National Science Foundation.



\bibliographystyle{IEEEtran}
\bibliography{IEEEabrv,bibliography}
%



\section{Appendix}
\label{sec:sec_ch5_appendix}
\renewcommand\thelemma{\unskip}
\renewcommand\thethm{\unskip}
\renewcommand\thecorollary{\unskip}
\subsection{Proof of Theorem~\ref{thm:thm_ch5_stat_dist}}
\label{subsec:subsec_ch5_proof_stationary_dist}
\begin{thm}
	The system described in Section~\ref{sec:sec_ch5_fcfs_low_priority} is stable, i.e. the average number of packets in the queue
	is finite, if and only if 
	\begin{equation}
		\label{eq:eq_ch5_stab_condition_2}
		\mu_1 > \lambda_1\lp 1+\frac{\lambda_2}{\mu_2}\rp.
	\end{equation}
	In this case the Markov chain shown in Fig.~\ref{fig:fig_ch5_sys_mc} has a stationary distribution $\Pi=[\pi_0,
	\pi_1,\dots,\pi_i,\dots,\pi'_1,\dots,\pi'_i,\dots]$, where $\pi_i$ denotes the stationary probability of state
	$q_i$, $i\geq 0$, and $\pi'_i$ denotes the stationary probability of state $q'_i$, $i>0$. This stationary
	distribution is described by the following system of equations,
	\begin{align}
		\pi_0 & = \frac{\mu_2}{\mu_2+\lambda_2}-\frac{\lambda_1}{\mu_1}, \label{eq:eq_ch5_po_2}\\
		\begin{bmatrix}
			\pi_i\\
			\pi'_i
		\end{bmatrix} &= \begin{bmatrix}\mathbf{\large{0}} & \mathbf{I}_2\end{bmatrix}\mathbf{H}^i\begin{bmatrix}
			\frac{\lambda}{\mu_1}-\frac{\mu_2\lambda_2}{\mu_1\lp\lambda_1+\mu_2\rp}\\
			\frac{\lambda_2}{\lambda_1+\mu_2}\\
			1\\
			0
		\end{bmatrix}\pi_0,\ i\geq1 \label{eq:eq_ch5_stat_prob_2}
	\end{align}
	where $\lambda=\lambda_1+\lambda_2$, $ 
		\mathbf{H} = \begin{bmatrix}
			\mathbf{C} & \mathbf{D}\\
			\mathbf{I}_2 & \mathbf{\large{0}}
		\end{bmatrix}$,
	\begin{align*}
		&\mathbf{C} = \begin{bmatrix}
			1+\frac{\lambda}{\mu_1}-\frac{\mu_2\lambda_2}{\mu_1\lp\mu_2+\lambda_1\rp}  &
			-\frac{\mu_2\lambda_1}{\mu_1\lp\mu_2+\lambda_1\rp}\\
			\frac{\lambda_2}{\mu_2+\lambda_1}  & \frac{\lambda_1}{\mu_2+\lambda_1}
		\end{bmatrix}, \mathbf{D} = \begin{bmatrix}
			-\frac{\lambda_1}{\mu_1} & 0\\
			0	& 	0
		\end{bmatrix}.
	\end{align*}
	$\mathbf{I}_2$ is the $2\times2$ identity matrix and $\mathbf{\large{0}}$ is the $2\times2$ zero matrix.
\end{thm}
\begin{proof}
Assume that 
	\begin{equation}
		\label{eq:eq_ch5_stab_condition_app}
		\mu_1 > \lambda_1\lp 1+\frac{\lambda_2}{\mu_2}\rp.
	\end{equation}
The detailed balance equations of the Markov chain given by Fig.~\ref{fig:fig_ch5_sys_mc} are given by:
\begin{equation}
\label{eq:eq_ch5_det_bal_eq}
\left\{\begin{aligned}
&\lambda\pi_0 = \mu_1\pi_1+\mu_2\pi'_1,\\
&(\lambda_1+\mu_2)\pi'_1 = \lambda_2\pi_0,\\
&\text{for $i\geq 1$,}\\
&\pi_{i+1} = \lp 1+\frac{\lambda}{\mu_1}-\frac{\mu_2\lambda_2}{\mu_1(\mu_2+\lambda_1)}\rp\pi_i
-\frac{\mu_2\lambda_1}{\mu_1(\mu_2+\lambda_1)}\pi'_i\\
&\qquad \quad -\frac{\lambda_1}{\mu_1}\pi_{i-1},\\
&\pi'_{i+1} = \frac{\lambda_2}{\mu_2+\lambda_1}\pi_i + \frac{\lambda_1}{\mu_2+\lambda_1}\pi'_i,
\end{aligned}\right.
\end{equation}
where $\lambda=\lambda_1+\lambda_2$.
For easier notation we denote
\begin{align*}
a_1 &=  1+\frac{\lambda}{\mu_1}-\frac{\mu_2\lambda_2}{\mu_1(\mu_2+\lambda_1)}, \hspace{\textwidth}\\
a_2 &= \frac{\mu_2\lambda_1}{\mu_1(\mu_2+\lambda_1)},\\
a_3 &= \frac{\lambda_1}{\mu_1},\\
a_4 &= \frac{\lambda_2}{\mu_2+\lambda_1},\\
a_5 &= \frac{\lambda_1}{\mu_2+\lambda_1}.
\end{align*}
Rewriting \eqref{eq:eq_ch5_det_bal_eq} in matrix form and using the above notation, we get
\begin{equation*}
\begin{bmatrix}
\pi_{i+1}\\
\pi'_{i+1}\\
\pi_i\\
\pi'_i
\end{bmatrix} = \begin{bmatrix}
a_1 & -a_2 & -a_3 & 0\\
a_4 & a_5 & 0 & 0\\
1 & 0 & 0 & 0\\
0 & 1 & 0 & 0
\end{bmatrix}\begin{bmatrix}
\pi_{i}\\
\pi'_{i}\\
\pi_{i-1}\\
\pi'_{i-1}
\end{bmatrix}.
\end{equation*}
Let $\mathbf{A}_i = \begin{bmatrix}
\pi_{i+1}\\
\pi'_{i+1}\\
\pi_i\\
\pi'_i
\end{bmatrix}$, $\mathbf{C} = \begin{bmatrix} a_1 & -a_2\\ a_4 & a_5\end{bmatrix}$, $\mathbf{D} = \begin{bmatrix} -a_3 & 0\\0 &
0\end{bmatrix}$ and $\mathbf{H}=\begin{bmatrix}\mathbf{C} &\mathbf{D} \\ \mathbf{I}_2 & \mathbf{0}\end{bmatrix}$. Then
\[ \mathbf{A}_i = \mathbf{H}\mathbf{A}_{i-1}.\]
Thus 
\begin{equation}
\label{eq:eq_ch5_A}
 \mathbf{A}_i = \mathbf{H}^i\mathbf{A}_0,\  i\geq 0
\end{equation}
where $\mathbf{A}_0 = \begin{bmatrix}
\pi_{1}\\
\pi'_{1}\\
\pi_0\\
0
\end{bmatrix}=\begin{bmatrix}\frac{\lambda}{\mu_1}-\frac{\mu_2\lambda_2}{\mu_1(\lambda_1+\mu_2)}\\
\frac{\lambda_2}{\lambda_1+\mu_2}\\ 1 \\ 0\end{bmatrix}\pi_0$, using the first two equations of system
\eqref{eq:eq_ch5_det_bal_eq}.

\eqref{eq:eq_ch5_A} shows that in order to find the stability criterion of the system in \eqref{eq:eq_ch5_det_bal_eq} we first need to
study the properties of $\mathbf{H}$. For that we compute its eigenvalues $l_0,l_1,l_2,l_3$ by solving the characteristic
equation $|l\mathbf{I}_4-\mathbf{H}|=0$. This leads to
\begin{equation}
\label{eq:eq_ch5_char_eq}
|l\mathbf{I}_4-\mathbf{H}| = l(l-1)(l^2-l(a_1+a_5-1)+a_3a_5).
\end{equation}
$\mathbf{H}$ has two obvious eigenvalues $l_0 = 0$ and $l_3 = 1$. To find the last two eigenvalues, let's find the root of the quadratic
polynomial 
\begin{equation}
\label{eq:eq_ch5_pol}
p(l) = l^2-l(a_1+a_5-1) + a_3a_5. 
\end{equation} 
It can be shown through simple algebra that the discriminant of the above polynomial is strictly positive. Hence the remaining
eigenvalues $l_1$ and $l_2$ are real and distinct. Let's assume that $l_1<l_2$. This means that the matrix $\mathbf{H}$ is diagonalizable and can be written
as $$\mathbf{H} = \mathbf{B}\mathbf{\Lambda}\mathbf{B}^{-1},$$ where the columns of $\mathbf{B}$ are the eigenvectors of
$\mathbf{H}$ and form a basis of $\mathbb{R}^4$. We denote by $\mathbf{e}_0, \mathbf{e}_1, \mathbf{e}_2, \mathbf{e}_3$ the eigenvectors corresponding to $l_0, l_1,
l_2, l_3$. 

So we can write $\mathbf{A}_0$ as 
\begin{equation}
\label{eq:eq_ch5_A0}
\mathbf{A}_0 = (\alpha_0 \mathbf{e}_0 + \alpha_1 \mathbf{e}_1 + \alpha_2 \mathbf{e}_2 + \alpha_3 \mathbf{e}_3)\pi_0,
\end{equation} 
with $\alpha_0, \alpha_1, \alpha_2, \alpha_3 \in \mathbb{R}$. Hence for $i>0$,
\begin{align}
\label{eq:eq_ch5_Ai}
\mathbf{A}_i &= \mathbf{H}^i\mathbf{A}_0 \nonumber\\
	     &= (\alpha_0\mathbf{H}^i \mathbf{e}_0 + \alpha_1\mathbf{H}^i \mathbf{e}_1 + \alpha_2\mathbf{H}^i \mathbf{e}_2 + \alpha_3\mathbf{H}^i
\mathbf{e}_3)\pi_0\nonumber\\
	     &= (\alpha_0 l_0^i \mathbf{e}_0 + \alpha_1 l_1^i \mathbf{e}_1 + \alpha_2 l_2^i \mathbf{e}_2 + \alpha_3 l_3^i \mathbf{e}_3)\pi_0\nonumber\\
	     &= (\alpha_1 l_1^i \mathbf{e}_1 + \alpha_2 l_2^i \mathbf{e}_2 + \alpha_3 \mathbf{e}_3)\pi_0,
\end{align} 
since $l_0=0$ and $l_3=1$. Equation~\eqref{eq:eq_ch5_Ai} shows that three conditions need to be satisfied for the system to be
stable and a steady-state distribution to exist:
\begin{itemize}
\item \textbf{Condition 1:}  $|l_1|<1$ and $|l_2|<1$.
\item \textbf{Condition 2:} $\alpha_3 = 0$.
\item \textbf{Condition 3:} $\alpha_1 l_1^i \mathbf{e}_1 + \alpha_2 l_2^i \mathbf{e}_2$ has positive components for all $i>0$.
\end{itemize}
\textbf{Condition 1} and \textbf{Condition 2} ensure that $$\lim_{i\to\infty} \pi_i = \lim_{i\to\infty} \pi'_i = 0$$ and thus the
sum of all probabilities, $\pi_0 + \sum_{i=1}^\infty (\pi_i+\pi'_i)$, does not diverge. \textbf{Condition 3} makes sure that the
components of $\mathbf{A}_i$ are positive probabilities. 
We will show that \eqref{eq:eq_ch5_stab_condition_app} is sufficient for the above three conditions to hold. 

Given that $l_1$ and $l_2$ are the roots of \eqref{eq:eq_ch5_pol} then the following holds
\begin{align}
\label{eq:eq_ch5_l1l2}
l_1l_2 &= a_3a_5\nonumber\\
l_1 + l_2 &= a_1 + a_5 -1.
\end{align} 
However, $l_1l_2 = a_3a_5 = \frac{\lambda_1^2}{\mu_1(\mu_2+\lambda_1)}\geq 0$. This means that either both $l_1$ and $l_2$ are
positive or they are both negative. Using \eqref{eq:eq_ch5_l1l2} again, we notice that $$ l_1+l_2 = a_1+a_5-1 =
\frac{\lambda_1\mu_2+\lambda_1^2+\lambda_1\lambda_2+\lambda_1\mu_1}{\mu_1(\mu_2+\lambda_1)}\geq 0.$$ This shows that both $l_1$ and $l_2$ are
strictly positive (since $0$ is not a root of $p(l)$). So to prove that \textbf{Condition 1} is satisfied we need to prove that $l_1<l_2<1$. This is equivalent to show that
\eqref{eq:eq_ch5_pol} evaluated at $1$ is strictly positive and that $l_1l_2<1$ since $p(l)$ is a convex quadratic
function in $l>0$. Using simple algebra it can be shown that  $$p(1) = 1-(a_1+a_5-1) + a_3a_5 = \frac{\mu_1\mu_2
-\lambda_1(\mu_2+\lambda_2)}{\mu_1(\mu_2+\lambda_1)}>0,$$ where the last inequality is due to \eqref{eq:eq_ch5_stab_condition_app}. Moreover,
\eqref{eq:eq_ch5_stab_condition_app} tells us that $\mu_1$ should be strictly bigger that $\lambda_1$. Thus we get that
$$l_1l_2 = \frac{\lambda_1}{\mu_1}\frac{\lambda_1}{\mu_2+\lambda_1}<1.$$ This shows that $0<l_1<l_2<1$ and that \textbf{Condition
1} is satisfied. 

To prove \textbf{Condition 2} we start by computing the eigenvectors of $\mathbf{H}$. For $l_0 = 0$, we solve the system given
by $\mathbf{H}\mathbf{e}_0 = \mathbf{0}$. If $\mathbf{e}_0 = \begin{bmatrix} u_1& u_2& u_3& 1\end{bmatrix}^T$ then \[ \begin{bmatrix}a_1 &-a_2
&-a_3& 0\\a_4 & a_5 & 0 & 0\\1 & 0& 0& 0\\0& 1& 0& 0 \end{bmatrix} \begin{bmatrix}u_1 \\ u_2\\ u_3\\
1\end{bmatrix}=\begin{bmatrix} 0\\ 0 \\ 0 \\ 0\end{bmatrix}.\] This system leads to $\mathbf{e}_0
=\begin{bmatrix}0& 0& 0& 1\end{bmatrix}^T$. Similarly, for $j=1,2,3$, if $\mathbf{e}_j=\begin{bmatrix}u_1 & u_2 & u_3 &
1\end{bmatrix}^T$ then solving the system 
\[ \begin{bmatrix}a_1 &-a_2
&-a_3& 0\\a_4 & a_5 & 0 & 0\\1 & 0& 0& 0\\0& 1& 0& 0 \end{bmatrix} \begin{bmatrix}u_1 \\ u_2\\ u_3\\
1\end{bmatrix}=l_j\begin{bmatrix} u_1\\ u_2 \\ u_3 \\ 1\end{bmatrix}\] leads to $\mathbf{e}_j=\begin{bmatrix}l_j(l_j-a_5)& l_ja_4&
l_j-a_5& a_4\end{bmatrix}^T$. 

We know that  $\mathbf{H}=\mathbf{B}\mathbf{\Lambda}\mathbf{B}^{-1}$. If \[\mathbf{\Lambda}=\begin{bmatrix}1 & 0 & 0 & 0\\ 0&
l_2& 0& 0\\ 0& 0& l_1& 0\\ 0& 0& 0& 0\end{bmatrix},\] then 
\[\mathbf{B} = \begin{bmatrix} 1-a_5 & l_2(l_2-a_5) & l_1(l_1-a_5)&0\\
			       a_4 & l_2 a_4& l_1 a_4& 0\\
			       1-a_5 & l_2-a_5 & l_1-a_5 & 0\\
			       a_4 & a_4 & a_4 & 1 \end{bmatrix}.\]
Note that the determinant of $\mathbf{B}$, $|\mathbf{B}|$, is non-zero when we assume \eqref{eq:eq_ch5_stab_condition_app}. Indeed, 
\[|\mathbf{B}| = a_4a_5(l_2-l_1)(-2+a_5-a_3a_5+a_1)<0\]
since $l_2>l_1$ and $-2+a_5-a_3a_5+a_1 =-p(1) <0$ as shown before. In order to compute $\alpha_3$, we rewrite \eqref{eq:eq_ch5_A0}
as follows
\[ \mathbf{A}_0 = \begin{bmatrix} \mathbf{e}_3& \mathbf{e}_2& \mathbf{e}_1& \mathbf{e}_0\end{bmatrix}\begin{bmatrix} \alpha_3\\ \alpha_2\\ \alpha_1\\
\alpha_0\end{bmatrix}\pi_0 = \mathbf{B}\begin{bmatrix} \alpha_3\\ \alpha_2\\ \alpha_1\\ \alpha_0\end{bmatrix}\pi_0.\]
But we also know that $$\mathbf{A}_0 = \begin{bmatrix}\frac{\lambda}{\mu_1}-\frac{\mu_2\lambda_2}{\mu_1(\lambda_1+\mu_2)}\\
\frac{\lambda_2}{\lambda_1+\mu_2}\\ 1 \\ 0\end{bmatrix}\pi_0=\begin{bmatrix}a_1 -1\\ a_4\\ 1\\ 0\end{bmatrix}\pi_0.$$ Thus 
\begin{equation}
\label{eq:eq_ch5_alpha3_sys}
\mathbf{B}\begin{bmatrix} \alpha_3\\ \alpha_2\\ \alpha_1\\ \alpha_0\end{bmatrix} =  \begin{bmatrix}a_1 -1\\ a_4\\ 1\\
0\end{bmatrix}.
\end{equation}
Solving the system in \eqref{eq:eq_ch5_alpha3_sys} with respect to $\alpha_3$, $\alpha_2$, $\alpha_1$ and $\alpha_0$ we get that
\begin{equation*}
	\begin{bmatrix}\alpha_3\\ \alpha_2\\ \alpha_1\\ \alpha_0\end{bmatrix} = \begin{bmatrix} 0\\ \frac{1}{l_2-l_1}
	\\ \frac{-1}{l_2-l_1}\\0\end{bmatrix}.
\end{equation*}
Thus $\alpha_3 = 0$ and \textbf{Condition 2} is proved. Note that we didn't need any assumptions to prove this condition.

Given the above results, we can now rewrite the system in \eqref{eq:eq_ch5_Ai} as 
\begin{equation}
	\label{eq:eq_ch5_final_sys}
	\left\{
		\begin{aligned}
			\mathbf{A}_i &= (\alpha_2 l_2^i\mathbf{e}_2 + \alpha_1 l_1^i\mathbf{e}_1)\pi_0,\qquad i>0\\
			\mathbf{A}_0 &= (\alpha_2\mathbf{e}_2 + \alpha_1\mathbf{e}_1)\pi_0 =\begin{bmatrix}a_1 -1\\ a_4\\ 1\\ 0\end{bmatrix}\pi_0.
		\end{aligned}
	\right.
\end{equation}
Using \eqref{eq:eq_ch5_final_sys} we can prove \textbf{Condition 3}. In fact, for any $i>0$,
\begin{align*}
	\alpha_2 l_2^i\mathbf{e}_2 + \alpha_1 l_1^i\mathbf{e}_1 &\overset{(a)}{=} \alpha_2\lp l_2^i\mathbf{e}_2 - l_1^i\mathbf{e}_1\rp\\
								&\overset{(b)}{\succ} \alpha_2 l_1^i(\mathbf{e}_2 -
								\mathbf{e}_1)\\
								&\overset{(c)}{=} l_1^i \begin{bmatrix}a_1 -1\\ a_4\\
									1\\ 0\end{bmatrix}\\
								&\overset{(d)}{\succ} \mathbf{0},
\end{align*}
where $\mathbf{x}\succ\mathbf{y}$ for some vectors $\mathbf{x}$ and $\mathbf{y}$ means that the components of
$\mathbf{x}-\mathbf{y}$ are strictly positive and 
\begin{itemize}
	\item[$(a)$] is because $\alpha_2=-\alpha_1$,
	\item[$(b)$] is because $0<l_1<l_2$,
	\item[$(c)$] is obtained from the second equality in \eqref{eq:eq_ch5_final_sys},
	\item[$(d)$] follows since $a_1-1 >0$ and $a_4>0$.
\end{itemize}

Up till now we have shown that if $\mu_1 >\lambda_1\lp 1+\frac{\lambda_2}{\mu_2}\rp$, the system described in
Section~\ref{sec:sec_ch5_system_model} is stable and a
steady-state distribution exists given by \eqref{eq:eq_ch5_final_sys}. The final point to prove in Theorem~\ref{thm:thm_ch5_stat_dist}
is the expression of $\pi_0$. For that we solve for $\pi_0$ the following equation
\begin{equation*}
	\pi_0 +\sum_{i=1}^{\infty} \pi_i+\pi'_i = \pi_0 + \begin{bmatrix}0 & 0& 1 &1\end{bmatrix}\sum_{i=1}^\infty \mathbf{A}_i
	=1.
\end{equation*}
Using the first equation of \eqref{eq:eq_ch5_final_sys} and replacing $\alpha_1$ and $\alpha_2$ by their expressions in function of
$l_1$ and $l_2$, using the fact that $l_1+l_2$ and $l_1 l_2$ are given by \eqref{eq:eq_ch5_l1l2} and finally replacing $a_1$, $a_2$, $a_3$, $a_4$ and $a_5$ by
their expressions in function of $\lambda_1$, $\lambda_2$, $\mu_1$, $\mu_2$ we get 
\[\pi_0 = \frac{\mu_2}{\mu_2+\lambda_2}-\frac{\lambda_1}{\mu_1}.\]
\end{proof}

\subsection{Proof of Corollary~\ref{cor:cor_ch5_exp_value}}
\label{appendix:appendix_proof_cor_ch5_exp_value}
\begin{corollary}
If we define $N(t)$ to be the number of stream $\mathcal{U}_1$ packets in the system at time $t$, then its moment generating
function is $\phi_{N(t)}$
\begin{equation}
	\resizebox{0.48\textwidth}{!}{$\phi_{N(t)}(s) = \pi_0\lp\frac{\mu_1\lp\lambda_1+\lambda_2+\mu_2-\lambda_1
	e^s\rp}{\mu_1\mu_2+\mu_1\lambda_1-e^s\lp\lambda_1^2+\lambda_1\lambda_2+\lambda_1\mu_1+\lambda_1\mu_2\rp+\lambda_1^2e^{2s}}\rp$},
\end{equation}
where $\pi_0$ is given by \eqref{eq:eq_ch5_po}. Particularly, the expected value of $N(t)$ is
\begin{equation}
\E\lp N(t)\rp = \frac{\lambda_1\lp 2\lambda_2\mu_2+\lambda_2\mu_1+\lambda_2^2+\mu_2^2\rp}{\lp \mu_2+\lambda_2\rp\lp
\mu_1\mu_2-\lambda_1\lp \mu_2+\lambda_2\rp\rp}.
\end{equation}
\end{corollary}
\begin{proof}

	At any point in time, there are exactly $i$ stream $\mathcal{U}_1$ packets in the system if we are in state $q_i$ or
	$q'_{i+1}$ in the Markov chain given by Fig.~\ref{fig:fig_ch5_sys_mc}. This means that the probability of having exactly $i$
	stream $\mathcal{U}_1$ packets in the system is $\pi_i+\pi'_{i+1}$. Hence, using the same quantities as in
	\Cref{subsec:subsec_ch5_proof_stationary_dist}
	\begin{align}
		\label{eq:eq_ch5_app_mgf_comp}
		\phi_{N(t)}(s) & \resizebox{0.42\textwidth}{!}{$
			       = \sum_{n=0}^\infty e^{sn}(\pi_i+\pi'_{i+1})
			       = \sum_{n=0}^\infty e^{sn}\lp\mathbf{A}_n^T\begin{bmatrix}0\\ 1\\
				       1\\ 0\end{bmatrix}\rp$}\nn
			       &= \sum_{n=0}^\infty e^{sn}\alpha_2\pi_0\lp (l_2^n\mathbf{e}_2-l_1^n\mathbf{e}_1)^T\begin{bmatrix}0\\ 1\\
												1\\ 0\end{bmatrix}\rp\nn
			       &= \resizebox{0.39\textwidth}{!}{$\alpha_2\pi_0\lp \sum_{n=0}^\infty
				\lp e^sl_2\rp^n\mathbf{e}_2^T\begin{bmatrix}0\\ 1\\
				 1\\ 0\end{bmatrix}-\sum_{n=0}^\infty \lp e^sl_1\rp^n\mathbf{e}_1^T\begin{bmatrix}0\\ 1\\
				 1\\ 0\end{bmatrix} \rp$}\nn
			 &= \resizebox{0.39\textwidth}{!}{$\alpha_2\pi_0\lp
				 \frac{1}{1-l_2e^s}(l_2a_4+l_2-a_5) -\frac{1}{1-l_1e^s}(l_1a_4+l_1-a_5)\rp$}\nn
			 &= \alpha_2\pi_0(l_2-l_1)\lp \frac{a_4+1-a_5e^s}{1-(l_1+l_2)e^s+l_1l_2e^{2s}}\rp.
		 \end{align}
		where the quantities used here are the one defined in the proof of Theorem~\ref{thm:thm_ch5_stat_dist}. Thus,
		 \begin{align*}
			 \resizebox{0.48\textwidth}{!}{$\phi_{N(t)}(s) =\pi_0\lp
			 \frac{\mu_1\lp \lambda_1+\lambda_2+\mu_2-\lambda_1
			 e^s\rp}{\mu_1\mu_2+\mu_1\lambda_1-e^s(\lambda_1\mu_2+\lambda_1^2+\lambda_1\lambda_2+\lambda_1\mu_1)+\lambda_1^2e^{2s}}\rp$}.
	\end{align*}
	This last equality is
	obtained by using \eqref{eq:eq_ch5_l1l2}, $\alpha_2 = \frac{1}{l_2-l_1}$ and replacing $a_1$, $a_2$, $a_3$, $a_4$,
	$a_5$ by their expressions in function of $\lambda_1$, $\lambda_2$, $\mu_1$ and $\mu_2$ in
	\eqref{eq:eq_ch5_app_mgf_comp}.
	Finally,
	\[ \resizebox{0.48\textwidth}{!}{$\E(N(t)) = \left.\frac{\mathrm{d}\phi_{N(t)}(s)}{\mathrm{d}s}\right|_{s=0} = \frac{\lambda_1\lp 2\lambda_2\mu_2+\lambda_2\mu_1+\lambda_2^2+\mu_2^2\rp}{\lp \mu_2+\lambda_2\rp\lp
	\mu_1\mu_2-\lambda_1\lp \mu_2+\lambda_2\rp\rp}$}.\]
\end{proof}

\subsection{Proof of Lemma~\ref{lemma:lemma_ch5_service_time_1} and overview on the detour flow graph method}
\label{appendix:appendix_proof_lemma_ch5_service_time_1}
\begin{lemma}
		Let $Y_j$ be the \lq\lq virtual\rq\rq\ service time of packet $j$ given that this packet does not
			find the system in state $q'_1$, i.e. $\Prob\lp Y_j>t\rp=\Prob\lp Z_j>t|\overline{\Psi_j}\rp$. Then,
			in steady state,
			\begin{equation}
				\label{eq:eq_ch5_mgf_Y_2}
				\phi_Y(s) = \E\lp e^{sY}\rp= \frac{\mu_1(\mu_2-s)}{s^2-s(\mu_2+\mu_1+\lambda_2)+\mu_1\mu_2}.
			\end{equation}
		 Similarly, let $Y'_j$ be the \lq\lq virtual\rq\rq\ service time of packet $j$ given that this packet 
			finds the system in state $q'_1$, i.e. $\Prob\lp Y'_j>t\rp=\Prob\lp Z_j>t|\Psi_j\rp$. Then, in steady
			state,
			\begin{equation}
				\label{eq:eq_ch5_mgf_Y'_2}
				\phi_{Y'}(s) = \E\lp e^{sY'}\rp= \frac{\mu_1\mu_2}{s^2-s(\mu_2+\mu_1+\lambda_2)+\mu_1\mu_2}.
			\end{equation}
\end{lemma}
\begin{proof}
	We start by proving \eqref{eq:eq_ch5_mgf_Y_2}. For this, we use the detour flow graph method. 
	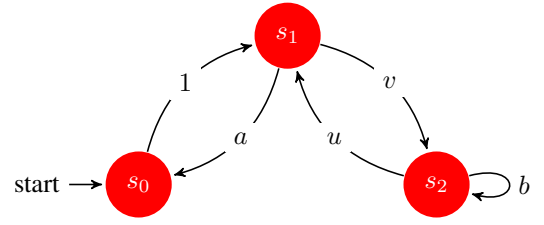
\begin{figure}[!t]
		\centering
	\begin{tikzpicture}[>=stealth',shorten >=1pt,auto,node distance=2.8cm, semithick]
		\tikzstyle{every state}=[fill=red,draw=none,text=white]
		\node[initial,state] (A)                    {$s_0$};
		\node[state]         (B) [above right of=A] {$s_1$};
		\node[state]         (C) [below right of=B] {$s_2$};
		
		\path [->] (A) edge [bend left]      node[ fill=white, anchor=center, pos=0.5] {$1$} (B);
		\path [->] (B) edge [bend left]      node[ fill=white, anchor=center, pos=0.5] {$a$} (A);
		\path [->] (B) edge [bend left]      node[ fill=white, anchor=center, pos=0.5] {$v$} (C);
		\path [->] (C) edge [loop right]     node {$b$} (C);
		\path [->] (C) edge [bend left]      node[ fill=white, anchor=center, pos=0.5] {$u$} (B);
	\end{tikzpicture}
	\caption{Semi-Markov chain representing the \lq\lq virtual\rq\rq\ service time $Y_j$.}
	\label{fig:fig_ch5_Y_mc}
	\end{figure}
	Fig.~\ref{fig:fig_ch5_Y_mc} shows the semi-Markov chain
	relative to the \lq\lq virtual\rq\rq\ service time $Y_j$ of the $j^{th}$ packet of first stream $\mathcal{U}_1$. Since
	the system is ergodic, it also applies to any packet at steady state. This
chain is constituted of three states: 
\begin{itemize}
\item $s_0$: in this state, the system is idle from a stream-$\str$ point of view. This means that no stream-$\str$ packet is being served. 
\item $s_1$: in this state, a stream-$\str$ packet is in service.
\item $s_2$: in this state, a stream-$\strr$ packet is in service.
\end{itemize}
	When the $j^{th}$ packet reaches the head of the buffer, the system is in the idle state $s_0$. Hence, with probability
	$1$ it goes immediately to state $s_1$ where it starts serving the $j^{th}$ packet. Due to the memoryless property 
	of the interarrival time of the second stream $X^{(2)}$, two clocks start: a service clock $S^{(1)}$ and a clock
	$X^{(2)}$. The service clock ticks first with probability $a=\Prob\lp S^{(1)}<X^{(2)}\rp$ and its value $A$ has
	distribution $\Prob\lp A>t\rp=\Prob\lp S^{(1)}>t|S^{(1)}<X^{(2)}\rp$. At this point, the stream $\mathcal{U}_1$ packet,
	currently being served, finishes service before any packet from the other stream is generated, and the system goes back
	to state $s_0$. This ends the \lq\lq virtual\rq\rq\ service time $Y_j$. Clock
	$X^{(2)}$ ticks first with probability $v=1-a=\Prob\lp X^{(2)}< S^{(1)}\rp$ and its value $V$ has distribution
	$\Prob\lp V>t\rp=\Prob\lp X^{(2)}>t|X^{(2)}< S^{(1)}\rp$. At this point, a new stream $\mathcal{U}_2$ update is
	generated and preempts the stream $\mathcal{U}_1$ packet currently in service. In this case, the system
	goes to state $s_2$, where the preempted stream $\mathcal{U}_1$ update is placed back at the head of the buffer, and the
	system starts service of the stream $\mathcal{U}_2$ update.  

	When the system arrives in state $s_2$, this means a new stream $\mathcal{U}_2$ packet was just generated and is
	starting its service. Thus, two clocks start: a service clock $S^{(2)}$ and a clock $X^{(2)}$. The service clock ticks
	first with probability $u=\Prob\lp S^{(2)}<X^{(2)}\rp$ and its value $U$ has distribution $\Prob\lp U>t\rp=\Prob\lp
	S^{(2)}>t|S^{(2)}< X^{(2)}\rp$. At this point, the packet currently being served finishes service before any new stream
	$\mathcal{U}_2$ packet is generated, and the system goes back to state $s_1$ where the $j^{th}$ packet of stream
	$\mathcal{U}_1$ starts its service again. However, clock $X^{(2)}$ ticks first with probability $b=1-u$, and
	its value $B$ has distribution $\Prob\lp B>t\rp=\Prob\lp X^{(2)}>t|X^{(2)}< S^{(2)}\rp$. At this point, a new stream
	$\mathcal{U}_2$ update is generated and preempts the one currently in service. In this case, the system stays in state
	$s_2$. 

	From the above analysis, we see that the \lq\lq virtual\rq\rq\ service time is given by the sum of the values of the different clocks
	on the path starting and finishing at $s_0$. For example, for the path $s_0s_1s_2s_1s_2s_2s_1s_0$ in
	Fig.~\ref{fig:fig_ch5_Y_mc}, the \lq\lq virtual\rq\rq\ service time $Y=V_1+U_1+V_2+B_1+U_2+A_1$, where all the random variables in the sum are
	mutually independent. This value of $Y$ is also valid for the path $s_0s_1s_2s_2s_1s_2s_1s_0$. Hence, $Y$ depends
	on the variables $A_j,B_j,U_j,V_j$ and their number of occurrences and not on the path itself. Therefore, the
	probability that exactly $(i_1, i_2, i_3, i_4)$ occurrences of $\lp A, B, U, V\rp$ occur, which is equivalent
	to the probability that 
	\begin{equation*}
		Y=\sum_{k=1}^{i_1} A_k + \sum_{k=1}^{i_2} B_k + \sum_{k=1}^{i_3} U_k + \sum_{k=1}^{i_4} V_k
	\end{equation*}
	is given by $a^{i_1}b^{i_2}u^{i_3}v^{i_4}Q(i_1,i_2,i_3,i_4)$, where 
	$Q(i_1,i_2,i_3,i_4)$ is the number of paths with this combination of occurrences. Taking into account the fact that
	the $\{A_k, B_k, U_k, V_k\}$ are mutually independent and denoting by $\{I_1, I_2, I_3, I_4\}$ the random variables
	associated with the number of occurrences of $\{A, B, U,V\}$ respectively, the moment generating function of $Y$ is,
	\begin{align}
		\label{eq:eq_ch5_Y_mgf1}
		\phi_Y(s) &= \E\lp\E\lp e^{sY}|\lp I_1, I_2, I_3, I_4\rp=\lp i_1, i_2, i_3, i_4\rp
		\rp\rp\nn
			  &= \sum_{i_1,i_2,i_3,i_4}
			  \left[a^{i_1}b^{i_2}u^{i_3}v^{i_4}Q(i_1,i_2,i_3,i_4)\right.\nn 
				  & \left. {} \E\lp e^{s\lp\sum_{k=1}^{i_1} A_k + \sum_{k=1}^{i_2} B_k + \sum_{k=1}^{i_3} U_k +
			  \sum_{k=1}^{i_4} V_k\rp}\rp\right]\nn
			  &=\sum_{i_1,i_2,i_3,i_4}
			  \left[a^{i_1}b^{i_2}u^{i_3}v^{i_4}Q(i_1,i_2,i_3,i_4)\right.\nn
				  &\left. {} \E\lp e^{sA}\rp^{i_1}\E\lp e^{sB}\rp^{i_2}\E\lp e^{sU}\rp^{i_3}\E\lp
			  e^{sV}\rp^{i_4}\right]. 
	\end{align}	
	
	In order to compute \eqref{eq:eq_ch5_Y_mgf1}, we modify the state diagram in Fig.~\ref{fig:fig_ch5_Y_mc} and represent
	it as a \emph{detour flow graph} (also called \emph{signal flow graph} \cite{mason1960,lincostello2004}) as shown in
	Fig.~\ref{fig:fig_ch5_detour_flow_graph_Y}. For that, we first notice that the ``virtual'' service time $Y$ is the
	interval of time spent by the system between two consecutive $s_0$ states. That's why, in
	Fig.~\ref{fig:fig_ch5_detour_flow_graph_Y}, we split the $s_0$ state into two states: a starting state $s_0$ and an end
	state $\bar{s}_0$. Hence, there is a one-to-one correspondence between the different paths from $s_0$ to $\bar{s}_0$
	and the different combinations in which we can write $Y$ in function of $(A, B, U,V)$. In order to capture the number
	of occurrences of the quantities $(A, B, U,V)$ over a certain path, we associate with each label four ``dummy''
	variables $(D_1,D_2,D_3,D_4)$ and the exponents of $(D_1,D_2,D_3,D_4)$ correspond to the number of occurrences of $(A,
	B, U,V)$ respectively. For example, if between two states $s_j$ and $s_i$ (for any $i,j$), the edge has a label that contains the
	factor $D_1D_4^2$, then it means that the system spent a time of $A+2V$ when passing from $s_j$ to $s_i$. Since we are
	interested in the distribution of $Y$, we multiply the labels of the edges in the detour flow graph by the probability
	of such label being visited. For example, given that the system is at state $s_1$, it jumps to state $s_2$ with
	probability $v$ and after spending a time $V$. Thus the label from $s_1$ to $s_2$ is $vD_4$. 

	Using Mason's gain formula \cite{mason1960}, we know that the generating function $H_1(D_1,D_2,D_3,D_4)$ of the detour flow graph
	shown in Fig.~\ref{fig:fig_ch5_detour_flow_graph_Y}, can be written as
	\begin{align}
		\label{eq:eq_ch5_gain_formula}
		&H_1(D_1,D_2,D_3,D_4)\nn
		&=\sum_{i_1,i_2,i_3,i_4}
			  \left[Q(i_1,i_2,i_3,i_4)a^{i_1}b^{i_2}u^{i_3}v^{i_4}
			   D_1^{i_1} D_2^{i_2}D_3^{i_3}D_4^{i_4}\right],
	\end{align}
	where $Q(i_1,i_2,i_3,i_4)$ is the number of paths with $i_1$ occurrences of $A$, $i_2$ occurrences of $B$,
	$i_3$ occurrences of $U$, $i_4$ occurrences of $V$.
	Comparing \eqref{eq:eq_ch5_Y_mgf1} and \eqref{eq:eq_ch5_gain_formula}, we notice that
	$$\phi_Y(s) = H_1\lp\E\lp e^{sA}\rp,\E\lp e^{sB}\rp,\E\lp e^{sU}\rp,\E\lp e^{sV}\rp\rp.$$
	
	Moreover, given a directed graph $G=(V,E)$ with algebraic label $L(e)$ on its
	edges, and a node $u\in V$ with no incoming edges, the transfer function
	$H(v)$ from $u$ to a node $v$ is the sum over of all paths from $u$ to
	$v$ with each path contributing the product of its edge labels to the
	sum (see \cite{mason1960,lincostello2004,rimoldi_2016}). The complete set of transfer functions $\{H(v): v\in V\}$ can be
	computed easily by solving the linear equations:
	$$\begin{cases}
		H(u) & = 1\\
		H(w) &= \sum_{w': (w',w)\in E} H(w') L(( w',w)), \quad w\neq u.
	\end{cases}$$
	Solving the system of linear equations above for the detour flow graph of Fig.~\ref{fig:fig_ch5_detour_flow_graph_Y},
	\eqref{eq:eq_ch5_gain_formula} becomes
	\begin{align}
		\label{eq:eq_ch5_H}
		H_1(D_1,D_2,D_3,D_4) &= \frac{aD_1(1-bD_2)}{1-bD_2-uD_3vD_4}.
	\end{align}
	From \cite[Appendix A, Lemma 2]{YatesKaul-2016arxiv}, we know that $A$, $B$, $U$ and $V$ are exponentially distributed
	with  $\E\lp e^{sB}\rp=\E\lp e^{sU}\rp=\frac{\lambda_2+\mu_2}{\lambda_2+\mu_2-s}$ and $\E\lp e^{sA}\rp=\E\lp
	e^{sV}\rp=\frac{\lambda_2+\mu_1}{\lambda_2+\mu_1-s}$. Simple computations show that 
	$a=\frac{\mu_1}{\mu_1+\lambda_2}$, $b=\frac{\lambda_2}{\mu_2+\lambda_2}$, $u=\frac{\mu_2}{\mu_2+\lambda_2}$,
	$v=\frac{\lambda_2}{\mu_1+\lambda_2}$. Finally, replacing the above
	expressions into \eqref{eq:eq_ch5_H}, we get our result.

	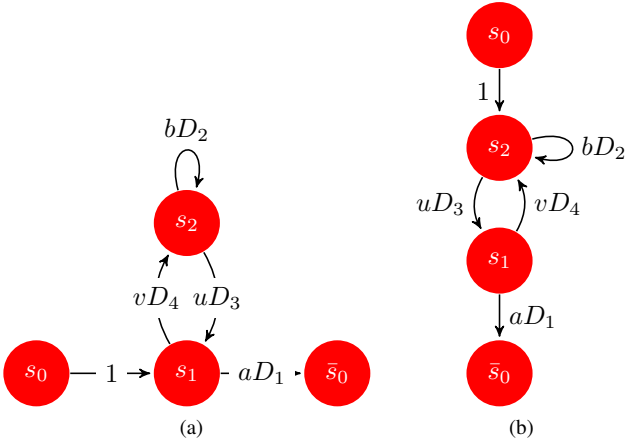
\begin{figure}[!t]
		\centering
		\subfloat[]{\begin{tikzpicture}[>=stealth',shorten >=1pt,auto,node distance=2cm, semithick,scale=0.6]
		\tikzstyle{every state}=[fill=red,draw=none,text=white]
		\node[state] (A)                     {$s_0$};
		\node[state] (B) [right of=A]  	     {$s_1$};
		\node[state] (C) [right of=B]  	     {$\bar{s}_0$};
		\node[state] (D) [above of=B]        {$s_2$};
		
		\path [->] (A) edge                  node[ fill=white, anchor=center, pos=0.5] {$1$} (B);
		\path [->] (B) edge   		     node[ fill=white, anchor=center, pos=0.5] {$aD_1$} (C);
		\path [->] (B) edge [bend left]	     node[ fill=white, anchor=center, pos=0.5] {$vD_4$} (D);
		\path [->] (D) edge [bend left]      node[ fill=white, anchor=center, pos=0.5] {$uD_3$} (B);
		\draw [->] (D) edge [loop above]     node {$bD_2$} (D);
	\end{tikzpicture}
	\label{fig:fig_ch5_detour_flow_graph_Y}}
	~
	\subfloat[]{
	\begin{tikzpicture}[>=stealth',shorten >=1pt,auto,node distance=1.5cm, semithick]
		\tikzstyle{every state}=[fill=red,draw=none,text=white]
		\node[state] (A)                     {$s_0$};
		\node[state] (B) [below of=A]  	     {$s_2$};
		\node[state] (C) [below of=B]  	     {$s_1$};
		\node[state] (D) [below of=C]        {$\bar{s}_0$};
		
		\path [->] (A) edge                  node[anchor=east] {$1$} (B);
		\path [->] (B) edge [bend right]     node[anchor=east] {$uD_3$} (C);
		\path [->] (B) edge [loop right]     node[anchor=west] {$bD_2$} (B);
		\path [->] (C) edge [bend right]     node[anchor=west] {$vD_4$} (B);
		\draw [->] (C) edge 		     node {$aD_1$} (D);
	\end{tikzpicture}
	\label{fig:fig_ch5_detour_flow_graph_Y'}}
	\caption{Detour flow graphs for (a) $Y$ and (b) $Y'$.}
	\label{fig:fig_ch5_detour_flow_graphs}
	\end{figure}

	To prove \eqref{eq:eq_ch5_mgf_Y'_2}, we use the same method as before. But in this case, we notice that the $j^{th}$ packet
	from stream $\mathcal{U}_1$ finds the system busy serving a packet from stream $\mathcal{U}_2$. This translates in the
	detour flow graph shown in Fig.~\ref{fig:fig_ch5_detour_flow_graph_Y'}. The generating function of this graph is 
	\begin{equation}
		\label{eq:eq_ch5_H_2}
		H_2(D_1,D_2,D_3,D_4) = \frac{aD_1uD_3}{1-bD_2-vD_4uD_3}.
	\end{equation}
	For {\small$\lp D_1,D_2,D_3,D_4\rp=\lp\E\lp e^{sA}\rp,\E\lp e^{sB}\rp,\E\lp e^{sU}\rp,\E\lp e^{sV}\rp\rp$} and replacing $a$,
	$b$, $u$ and $v$ by their values in \eqref{eq:eq_ch5_H_2}, we obtain \eqref{eq:eq_ch5_mgf_Y'_2}.
\end{proof}

\subsection{Proof of Lemma~\ref{lemma:lemma_ch5_system_time_lb}}
\label{appendix:appendix_proof_lemma_ch5_system_time_lb}
\begin{lemma}
	Assume an M/G/1 queue with interarrival time $X^{(1)}$ exponentially distributed with rate $\lambda_1$ and service time
	$Y$ whose moment generating function is given by \eqref{eq:eq_ch5_mgf_Y}. The service time and the interarrival time are
	assumed to be independent. Then the distribution of the system time $T$ is 
	{\small\begin{equation}
		f_T(t) = C_1 e^{-\alpha_1 t}(\mu_2-\alpha_1) - C_1 e^{-\alpha_2 t}(\mu_2-\alpha_2),\ t\geq0,
	\end{equation}}%
	where $\alpha_1,\alpha_2>0$ are the roots of the quadratic expression
	$$s^2-s(\mu_1+\mu_2+\lambda_2-\lambda_1)+\mu_1\mu_2-\lambda_1\mu_2-\lambda_1\lambda_2,$$
	$$C_1 = \frac{(1-\rho)\mu_1}{\alpha_2-\alpha_1},$$
	and $\rho = \lambda_1\E\lp Y\rp =
	\frac{\lambda_1(\mu_2+\lambda_2)}{\mu_1\mu_2}$.
\end{lemma}
\begin{proof}
From \cite[p. 166]{daigle2005queueing}, we know that the Laplace transform of the system time $T$ is 
$$ \E\lp e^{-sT}\rp = \frac{(1-\rho)s\phi_Y(-s)}{s-\lambda_1(1-\phi_Y(-s))}.$$ Replacing $\phi_Y(-s)$ by its expression in \eqref{eq:eq_ch5_mgf_Y_2} we
get 
\begin{align}
\label{eq:eq_ch5_syst_time_LT1}
	\E\lp e^{-sT}\rp &=
	\resizebox{0.38\textwidth}{!}{$\frac{(1-\rho)\mu_1(\mu_2+s)}{s^2+s(\mu_1+\mu_2+\lambda_2-\lambda_1)+\mu_1\mu_2-\lambda_1\mu_2-\lambda_1\lambda_2}$}\nn
	          &= \frac{(1-\rho)\mu_1(\mu_2+s)}{(s-s_1)(s-s_2)}\nn
		  &= s\frac{(1-\rho)\mu_1}{(s-s_1)(s-s_2)} + \frac{(1-\rho)\mu_1\mu_2}{(s-s_1)(s-s_2)},
\end{align}
where $s_1$ and $s_2$ are two real roots of the quadratic equation
\[ s^2+s(\mu_1+\mu_2+\lambda_2-\lambda_1)+\mu_1\mu_2-\lambda_1\mu_2-\lambda_1\lambda_2.\]
Moreover, due to condition \eqref{eq:eq_ch5_stab_condition}, $$s_1+s_2 =
-\mu_1-\mu_2-\lambda_2+\lambda_1<0$$ and $$s_1s_2= \mu_1\mu_2-\lambda_1\mu_2-\lambda_1\lambda_2>0.$$ This proves
that both roots $s_1$ and $s_2$ are negative. Let
\[ G(s) = \frac{(1-\rho)\mu_1}{(s-s_1)(s-s_2)},\]
and $g(t)$ its inverse Laplace transform. Using the initial value theorem:
\begin{align}
\label{eq:eq_ch5_sys_time_ivt}
g(0^{+}) = \lim_{s\to\infty} sG(s) = 0.
\end{align}
Using \eqref{eq:eq_ch5_sys_time_ivt} and the expression of $G(s)$, \eqref{eq:eq_ch5_syst_time_LT1} can be written as
\begin{align}
\label{eq:eq_ch5_syst_time_LT2}
\E\lp e^{-sT}\rp &= sG(s) - g(0^{+}) +\mu_2G(s).
\end{align}
Therefore, the probability density function of the system time $f_T(t)$ (which is the inverse Laplace transform of $\E\lp
e^{-sT}\rp$) is 
\begin{align}
\label{eq:eq_ch5_syst_time_ilt}
f_T(t) = g'(t) + \mu_2 g(t).
\end{align}
By partial fraction expansion, \[ G(s) = \frac{C_1}{s-s_1}-\frac{C_1}{s-s_2},\] where $C_1 =
\frac{(1-\rho)\mu_1}{s_1-s_2}$. Denoting $\alpha_1 = -s_1 >0$ and $\alpha_2= -s_2>0$, we get 
\[ G(s) = \frac{C_1}{s+\alpha_1}-\frac{C_1}{s+\alpha_2},\text{ and } C_1 = \frac{(1-\rho)\mu_1}{\alpha_2-\alpha_1}.\]
Thus,
\[g(t) = C_1 e^{-\alpha_1t}-C_1 e^{-\alpha_2t},\]
and 
\[ f_T(t) = C_1 e^{-\alpha_1t}(\mu_2-\alpha_1)-C_1 e^{-\alpha_2t}(\mu_2-\alpha_2).\]
\end{proof}
\subsection{Proof of Lemma~\ref{lemma:lemma_ch5_low_preempt_mg11_Y}}
\label{appendix:appendix_proof_lemma_ch5_low_preempt_mg11_Y}
\begin{lemma}
	The moment generating function of the interdeparture time of stream $\mathcal{U}_1$, $Y$, is 
	\begin{equation}
		\label{eq:eq_ch5_low_preempt_mg11_Y_2}
		\phi_{Y}(s) = \frac{\lambda_1 P_{\lambda-s}\lp \lambda_2 L_{\lambda_2-s}-s\rp}{\lambda_1 P_{\lambda-s}\lp
\lambda_2 L_{\lambda_2-s} -s\rp-s(\lambda_2-s)}. 
	\end{equation}
\end{lemma}

\begin{proof}
	We use the detour flow graph method. 	
	\begin{figure}[t]
		\centering
	\begin{tikzpicture}[>=stealth',shorten >=1pt,auto,node distance=2.8cm, semithick]
		\tikzstyle{every state}=[fill=red,draw=none,text=white]
		\node[initial,state] (A)                    {$q_0$};
		\node[state]         (B) [above right of=A] {$q_1$};
		\node[state]         (D) [below right of=A] {$q_{0'}$};
		\node[state]         (C) [below right of=B] {$q_{1'}$};
		
		\path [->] (A) edge [bend left]      node[ fill=white, anchor=center, pos=0.5] {$a$} (B);
		\path [->] (A) edge                  node[ fill=white, anchor=center, pos=0.5] {$z$} (C);
		\path [->] (B) edge [loop above]     node {$b$}(B);
		\path [->] (B) edge [bend left]      node[ fill=white, anchor=center, pos=0.5] {$u$} (A);
		\path [->] (B) edge [bend left]      node[ fill=white, anchor=center, pos=0.5] {$d$} (C);
		\path [->] (C) edge [bend left]      node[ fill=white, anchor=center, pos=0.5] {$f$} (D);
		\path [->] (C) edge [loop right]     node {$v$} (C);
		\path [->] (D) edge [bend left]      node[ fill=white, anchor=center, pos=0.5] {$z$} (C);
		\draw [->] (D) ..controls +(east:5) and +(east:5)..(B) node[ fill=white, anchor=center, pos=0.5] {$a$};
	\end{tikzpicture}
	\caption{Semi-Markov chain representing the M/G/1/1 interdeparture time for stream $\str$.}
	\label{fig:fig_ch5_low_preempt_Y_mc}
	\end{figure}
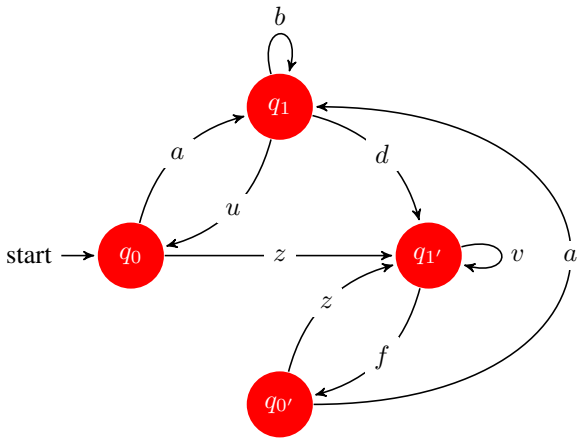
	We define $\Lambda=\min\lp X^{(1)},X^{(2)}\rp$. As $\Lambda$ is the minimum of
	independent exponential random variables, then it is also exponentially distributed with rates
	$\lambda=\lambda_1+\lambda_2$. Fig.~\ref{fig:fig_ch5_low_preempt_Y_mc} shows the semi-Markov chain
	relative to the interdeparture time $Y_j$ between the $j^{th}$  and $j+1^{th}$ successfully received packet of stream
$\mathcal{U}_1$. This chain is composed of 4 states:
\begin{itemize}
\item $q_0$ is the idle state reached after the reception of a stream-$\str$ packet. This means the system is empty. 
\item $q_1$ is the state where a stream-$\str$ packet is being served.
\item $q_{1'}$ is the state where a stream-$\strr$ packet is being served.
\item $q_{0'}$ is the idle state reached after the reception of a stream-$\strr$ packet. At this point also the system is
	empty. The need to differentiate between states $q_0$ and $q_{0'}$ will become clear shortly after. 
\end{itemize}
	When the $j^{th}$ packet is delivered to the monitor, the system is in the idle state $q_0$. Due to the memoryless
	property of the interarrival times of both streams, two clocks start: a clock $X^{(1)}$ and a clock $X^{(2)}$. Clock $X^{(1)}$
	ticks first with probability $a=\Prob\lp X^{(1)}<X^{(2)}\rp$, at which point a new packet from stream $\str$ will be generated
	first and the system goes to state $q_1$. The value $A$ of the clock when it ticks has distribution $\Prob\lp A>t\rp=\Prob\lp
	X^{(1)}>t|X^{(1)}<X^{(2)}\rp$. Clock $X^{(2)}$ ticks first with probability $z=1-a=\Prob\lp X^{(2)}<X^{(1)}\rp$, at which point
	a new packet from stream $\mathcal{U}_2$ is generated first and the system goes to state $q_{1'}$. The value
	$Z$ of this second clock when it ticks has distribution $\Prob\lp Z>t\rp=\Prob\lp X^{(2)}>t|X^{(2)}<X^{(1)}\rp$.

	When the system arrives in state $q_1$, this means a packet from stream $\str$ is starting its service. Thus, due to the memoryless property 
	of $X^{(2)}$, three clocks start: a service clock $S^{(1)}$, clock $X^{(1)}$ and clock $X^{(2)}$. The service clock ticks
	first with probability $u=\Prob\lp S^{(1)}<\Lambda\rp$ and its value $U$ has distribution $\Prob\lp U>t\rp=\Prob\lp
	S^{(1)}>t|S^{(1)}<\Lambda\rp$. At this point, the stream $\str$ packet currently being served finishes service before any new packet is
	generated and the system goes back to state $q_0$. This ends the interdeparture time $Y_j$. Clock
	$X^{(1)}$ ticks first with probability $b=\Prob\lp X^{(1)}<\min\lp S^{(1)},X^{(2)}\rp\rp$ and its value $B$ has distribution
	$\Prob\lp B>t\rp=\Prob\lp X^{(1)}>t|X^{(1)}<\min\lp S^{(1)},X^{(2)}\rp\rp$. At this point, a new stream $\str$ update is
	generated before any other update from other streams and preempts the one currently in service. In this case the system
	stays in state $q_1$. The third clock $X^{(2)}$ ticks first with probability $d=\Prob\lp X^{(2)}<\min\lp
	S^{(1)},X^{(1)}\rp\rp$ and its value $D$ has distribution $\Prob\lp D>t\rp=\Prob\lp X^{(2)}>t|X^{(2)}<\min\lp
	S^{(1)},X^{(1)}\rp\rp$. At this point, a new update from stream $\strr$ is generated, preempts the one currently in service
	and the system switches to state $q_{1'}$. 

	When the system arrives in state $q_{1'}$, this means a packet from stream $\strr$ is starting its service. Thus, due to the memoryless property 
	of $X^{(2)}$, two clocks are of interest: a service clock $\SsS$ and clock $\XX$. What happens to stream $\str$ is
irrelevant, as it has lower priority and any generated packet will be discarded. The service clock ticks
	first with probability $f = \Prob\lp \SsS < \XX\rp$  and its value $F$ is
distributed according to $\Prob\lp F>t\rp = \Prob\lp \SsS>t|\SsS<\XX\rp$. At this point, the stream $\strr$ packet currently being served finishes service before any new packet is
	generated and the system goes to state $q_{0'}$. Otherwise, clock $\XX$ ticks first with
	probability $v=1-f = \Prob\lp \XX<\SsS \rp$ and has value $V$ distributed as $\Prob \lp V>t\rp=\Prob\lp \XX>t|\XX<\SsS
\rp$. At this point, a new update from stream $\strr$ is generated, preempts the one currently in service
	and the system stays in state $q_{1'}$.

	Finally, when the system arrives in state $q_{0'}$, this means the system is idle but no update from stream $\str$ has
	been delivered. Given that $X^{(1)}$ and $\XX$ are memoryless, the system in state $q_{0'}$ behaves exactly as if it
	were in state $q_0$.

	From the above analysis, we see that the interdeparture time is given by the sum of the values of the different clocks
	on the path starting and finishing at $q_0$. For example, for the path $q_0q_1q_{1'}q_{0'}q_{1'}q_{0'}q_1q_0$ in
	Fig.~\ref{fig:fig_ch5_low_preempt_Y_mc}, the interdeparture time $Y=A_1+D_1+F_1+Z_1+F_2+A_2+U_1$, where all the random variables in the sum are
	mutually independent. This value of $Y$ is also valid for the path $q_0q_{1'}q_{0'}q_1q_{1'}q_{0'}q_1q_0$. Hence $Y$ depends
	on the variables $A_j,B_j,D_j,F_j,U_j,V_j,Z_j$ and their number of occurrences and not on the path itself. Therefore, the
	probability that exactly $(i_1, i_2, i_3, i_4, i_5, i_6, i_7)$ occurrences of $\lp A, B, D, F, U, V, Z\rp$ happen, which is equivalent
	to the probability that 	
	\[\resizebox{0.48\textwidth}{2.5ex}{$Y=\sum_{k=1}^{i_1} A_k + \sum_{k=1}^{i_2} B_k + \sum_{k=1}^{i_3} D_k + \sum_{k=1}^{i_4} F_k +
	\sum_{k=1}^{i_5} U_k + \sum_{k=1}^{i_6} V_k +
	\sum_{k=1}^{i_7} Z_k$}\]
	is given by $a^{i_1}b^{i_2}d^{i_3}f^{i_4}u^{i_5}v^{i_6}z^{i_7}Q(i_1,i_2,i_3,i_4,i_5,i_6,i_7)$, where 
	$Q(i_1,i_2,i_3,i_4,i_5,i_6,i_7)$ is the number of paths with this combination of occurrences. Taking into account the fact that
	the $\{A_k, B_k,D_k, F_k, U_k, V_k, Z_k\}$ are mutually independent, the moment generating function of $Y$ is	
	\begin{align}
		\label{eq:eq_ch5_low_preempt_mg11_Y_2_mgf1}
		\phi_Y(s) &= \resizebox{0.41\textwidth}{!}{$\E\lp\E\lp e^{sY}|\lp I_1, I_2, I_3, I_4, I_5, I_6, I_7\rp=\lp i_1, i_2, i_3, i_4, i_5, i_6, i_7\rp
	\rp\rp$}\nn
	&= \resizebox{0.41\textwidth}{!}{$\sum_{\substack{i_1,i_2,i_3,\\i_4,i_5,i_6,i_7}}
	\left[a^{i_1}b^{i_2}d^{i_3}f^{i_4}u^{i_5}v^{i_6}z^{i_7}Q(i_1,i_2,i_3,i_4,i_5,i_6,i_7)\right.$}\nn 
		& \resizebox{0.43\textwidth}{2.5ex}{$\left.  \E\lp e^{s\lp\sum_{k=1}^{i_1} A_k + \sum_{k=1}^{i_2} B_k + \sum_{k=1}^{i_3} D_k +
			  \sum_{k=1}^{i_4} F_k +\sum_{k=1}^{i_5}U_k + \sum_{k=1}^{i_6} V_k + \sum_{k=1}^{i_7}
		  Z_k\rp}\rp\right]$}\nn
		  &=\resizebox{0.41\textwidth}{!}{$\sum_{\substack{i_1,i_2,i_3,\\i_4,i_5,i_6,i_7}}
		  \left[a^{i_1}b^{i_2}d^{i_3}f^{i_4}u^{i_5}v^{i_6}z^{i_7}Q(i_1,i_2,i_3,i_4,i_5,i_6,i_7)\right.$}\nn
			  &\resizebox{0.43\textwidth}{!}{$\left. {} \E\lp e^{sA}\rp^{i_1}\E\lp e^{sB}\rp^{i_2}\E\lp e^{sD}\rp^{i_3}\E\lp
		  e^{sF}\rp^{i_4}\E\lp e^{sU}\rp^{i_5}\E\lp e^{sV}\rp^{i_6}\E\lp e^{sZ}\rp^{i_7}\right]$},
	\end{align}
	where $\{I_1, I_2, I_3, I_4, I_5, I_6, I_7\}$ are the random variables associated with the number of occurrences of
$\{A, B, D,F, U,
	V, Z\}$, respectively.
	
	\begin{figure}[!t]
		\centering
	\begin{tikzpicture}[>=stealth',shorten >=1pt,auto,node distance=2.6cm, semithick]
		\tikzstyle{every state}=[fill=red,draw=none,text=white]
		\node[state] (A)                     {$q_0$};
		\node[state] (C) [right of=A]  	     {$q_{1'}$};
		\node[state] (B) [right of=C]  	     {$q_1$};
		\node[state] (E) [right of=B] 	     {$\bar q_{0}$};
		\node[state] (D) at (3.9,2.5)  	     {$q_{0'}$};
		
		\path [->] (A) edge [bend right]     node[ fill=white, anchor=center, pos=0.5] {$aW_1$} (B);
		\path [->] (A) edge                  node[ fill=white, anchor=center, pos=0.5] {$zW_7$} (C);
		\draw [->] (B) to [out=60, in=30, looseness=8] node[above] {$bW_2$} (B);
		\path [->] (B) edge                  node[ fill=white, anchor=center, pos=0.5] {$uW_5$} (E);
		\path [->] (B) edge 		     node[ fill=white, anchor=center, pos=0.5] {$dW_3$} (C);
		\path [->] (C) edge [bend left=30]   node[ fill=white, anchor=center, pos=0.5] {$fW_4$} (D);
		\draw [->] (C) to [out=120, in=150, looseness=8] node[above] {$vW_6$}(C);
		\path [->] (D) edge [bend left=20]   node[ fill=white, anchor=center, pos=0.5] {$zW_7$} (C);
		\draw [->] (D) edge                  node[ fill=white, anchor=center, pos=0.5] {$aW_1$} (B);
	\end{tikzpicture}
	\caption{Detour flow graph of the M/G/1/1 interdeparture time for stream $\str$.}
	\label{fig:fig_ch5_low_preempt_detour_flow_graph}
	\end{figure}
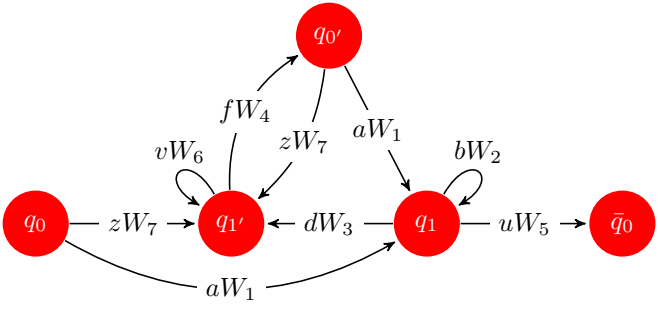
	In order to compute \eqref{eq:eq_ch5_low_preempt_mg11_Y_2_mgf1}, we modify the state diagram in
	Fig.~\ref{fig:fig_ch5_low_preempt_Y_mc} and represent
	it as a \emph{detour flow graph} (also called \emph{signal flow graph} \cite{mason1960,lincostello2004}) as shown in
	Fig.~\ref{fig:fig_ch5_low_preempt_detour_flow_graph}. For that, we first notice that the interdeparture time $Y$ is the
	interval of time spent by the system between two consecutive $q_0$ states. That's why, in
	Fig.~\ref{fig:fig_ch5_low_preempt_detour_flow_graph}, we split the $q_0$ state into two states: a starting state $q_0$ and an end
	state $\bar{q}_0$. Hence, there is a one-to-one correspondence between the different paths from $q_0$ to $\bar{q}_0$
	and the different combinations in which we can write $Y$ in function of $(A, B, D, F,U,V,Z)$. In order to capture the number
	of occurrences of the quantities $(A, B, D,F,U,V,Z)$ over a certain path, we associate with each label seven ``dummy''
	variables $(W_1,W_2,W_3,W_4,W_5,W_6,W_7)$ and the exponents of $(W_1,W_2,W_3,W_4,W_5,W_6,W_7)$ correspond to the number
	of occurrences of $(A, B, D, F,U,V,Z)$ respectively. For example, if between two states $q_j$ and $q_i$ (for any
	$i,j$), the edge has a label that contains the	factor $W_1W_3^2$, then it means that the system spent a time of $A+2D$
	when passing from $q_j$ to $q_i$. Since we are interested in the distribution of $Y$, we multiply the labels of the
	edges in the detour flow graph by the probability of such label being visited. For example, given that the system is at
	state $q_{1'}$, it jumps to state $q_{0'}$ with probability $f$ and after spending a time $F$. Thus the label from
	$q_{1'}$ to $q_{0'}$ is $fW_4$. 

	Using Mason's gain formula \cite{mason1960}, we know that the generating function $H(W_1,W_2,W_3,W_4,W_5,W_6,W_7)$ of the detour flow graph
	shown in Fig.~\ref{fig:fig_ch5_low_preempt_detour_flow_graph}, can be written as
	\begin{align}
		\label{eq:eq_ch5_gain_formula_2}
		&H(W_1,W_2,W_3,W_4,W_5,W_6,W_7)\nn
		&=\sum_{\substack{i_1,i_2,i_3,\\i_4,i_5,i_6,i_7}}
			  \left[Q(i_1,i_2,i_3,i_4,i_5,i_6,i_7)a^{i_1}b^{i_2}d^{i_3}f^{i_4}u^{i_5}v^{i_6}z^{i_7}\right.\nn
			  &\qquad \left. {} W_1^{i_1} W_2^{i_2}W_3^{i_3}W_4^{i_4}W_5^{i_5}W_6^{i_6}W_7^{i_7}\right],
	\end{align}
	where $Q(i_1,i_2,i_3,i_4,i_5,i_6,i_7)$ is the number of paths with $i_1$ occurrences of $A$, $i_2$ occurrences of $B$,
	$i_3$ occurrences of $D$, $i_4$ occurrences of $F$, $i_5$ occurrences of $U$, $i_6$ occurrences of $V$, $i_7$
	occurrences of $Z$.
	Comparing \eqref{eq:eq_ch5_low_preempt_mg11_Y_2_mgf1} and \eqref{eq:eq_ch5_gain_formula_2}, we notice that
	\begin{align*}
		\phi_Y(s) &= \resizebox{0.4\textwidth}{2ex}{$H\lp\E\lp e^{sA}\rp,\E\lp e^{sB}\rp,\E\lp e^{sD}\rp, \E\lp
		e^{sF}\rp,\E\lp e^{sU}\rp,\E\lp e^{sV}\rp,\E\lp e^{sZ}\rp\rp$}.
	\end{align*}
	Moreover, given a directed graph $G=(V,E)$ with algebraic label $L(e)$ on its
	edges, and a node $u\in V$ with no incoming edges, the transfer function
	$H(v)$ from $u$ to a node $v$ is the sum over of all paths from $u$ to
	$v$ with each path contributing the product of its edge labels to the
	sum (see \cite{mason1960,lincostello2004,rimoldi_2016}). The complete set of transfer functions $\{H(v): v\in V\}$ can be
	computed easily by solving the linear equations:
	$$\begin{cases}
		H(u) & = 1\\
		H(w) &= \sum_{w': (w',w)\in E} H(w') L(( w',w)), \quad w\neq u.
	\end{cases}$$
	Solving the system of linear equations above yields the transfer function as
	\begin{align}
		\label{eq:eq_ch5_low_preempt_H}
		&H(W_1,W_2,W_3,W_4,W_5,W_6,W_7) \nn
		&=\frac{uW_5aW_1(1-vW_6)}{\lp 1-zW_7fW_4-vW_6\rp\lp 1-bW_2\rp - dW_3aW_1fW_4}.
	\end{align}
		Using \cite[Lemma1]{NajmTelatar2018} and Lemma~\ref{lemma:lemma_ch5_low_preempt_mg11_sys_time}, we know that $\E\lp e^{sB}\rp=\E\lp
	e^{sD}\rp=\frac{\lambda\lp1-P_{\lambda-s}\rp}{\lp\lambda-s\rp\lp1-P_\lambda\rp}$, $\E\lp e^{sA}\rp=\E\lp
	e^{sZ}\rp=\frac{\lambda}{\lambda-s}$, $\E\lp e^{sF}\rp = \frac{L_{\lambda_2-s}}{L_{\lambda_2}}$ and $\E\lp e^{sV}\rp =
\frac{\lambda_2(1-L_{\lambda_2-s})}{(\lambda_2-s)(1-L_{\lambda_2})}$. Moreover, we can notice that $U$ has the same distribution as the system time
	$T$ so $\E\lp e^{sU}\rp=\frac{P_{\lambda-s}}{P_\lambda}$. Simple computations show that 
	$a=\frac{\lambda_1}{\lambda}$, $b=\frac{\lambda_1}{\lambda}\lp1-P_\lambda \rp$, 
	$d=\frac{\lambda_2}{\lambda}\lp1-P_\lambda\rp$, $f=L_{\lambda_2}$, $u=P_\lambda$, $v=1-L_{\lambda_2}$, $z=\frac{\lambda-\lambda_1}{\lambda}$. Finally, replacing the above
	expressions into \eqref{eq:eq_ch5_low_preempt_H}, we get our result.
\end{proof}

\subsection{Proof of Lemma~\ref{lemma:lemma_ch3_T_Y_independ}}
\label{appendix:appendix_proof_lemma_ch3_T_Y_independ}
\begin{lemma}
	Consider stream $\str$. For any $j\geq1$, the random variables $T_j$ and $Y_j$ relative to the $j^{th}$ successful packet are independent.
	Moreover the process $(Y_j)_{j\geq1}$ is i.i.d, with its distribution given by
\Cref{lemma:lemma_ch5_low_preempt_mg11_Y}, and the process $R(\tau)=\sup\{n\in\mathbb{N};D_n\leq\tau\}$ is a renewal process. 
\end{lemma}
\begin{proof}
Let $L_j=\min\lp X^{(1)}_j,X^{(2)}\rp$. Since the interarrival times for both streams are exponential and independent, $L_j$ is also
exponential with rate $\lambda=\lambda_1+\lambda_2$. Except $L_j$, all other variables are relative to stream $\str$. The
$j^{th}$ successful packet leaves the queue empty hence
$Y_j=\hat{X}_j+Z_j$. $\hat{X}_j=L_j-T_j$ is the remaining time between the departure of the stream-$\str$ $j^{th}$
successful packet, and the generation time of the next packet to be transmitted (it can belong to stream $\str$ or stream
$\strr$). $Z_j$ is the time for a new stream-$\str$ packet to be successfully delivered. $Z_j$ does not overlap with $T_j$ and
thus is independent from it. As for $\hat{X}_j$, we also obtain that it is independent of $T_j$. To prove this,  notice that
for a successfully received packet $j$ the joint distribution $f_{L_j,T_j}(x,t)$ can be written as
\begin{equation}
 \label{eq:eq_ch3_eq11}
 f_{L_j,T_j}(x,t) = f_{L,T|L\geq T}(x,t|x\geq t) = \left\{\begin{array}{ll}
                       0 & \text{if\ }x<t\\
                       \frac{f_{L,S}(x,t)}{\mathbb{P}(S<L)} & \text{if\ }x>t
                      \end{array}\right.,
\end{equation}
where $L=\min\lp X^{(1)},X^{(2)}\rp$ and $S$ is the generic service time. These two variables are independent. Now, using a change of variable
we obtain
\begin{align}
 \label{eq:eq_ch3_eq12}
 f_{\hat{X}_j,T_j}(\hat{x},t) &= f_{L_j-T_j,T_j}(\hat{x},t)= f_{L_j,T_j}(\hat{x}+t,t)\nonumber\\
			      &= \left\{\begin{array}{ll}
				      0 & \text{if\ }\hat{x}<0\\
		       \frac{f_{L,S}(\hat{x}+t,t)}{\mathbb{P}(S<L)} & \text{if\ }\hat{x}>0
                      \end{array}\right.\nonumber\\
			      &= \left\{\begin{array}{ll}
				      0 & \text{if\ }\hat{x}<0\\
				      \frac{\lambda e^{-\lambda(\hat{x}+t)}f_S(t)}{\mathbb{P}(S<L)} & \text{if\ }\hat{x}>0
			      \end{array}\right.\nonumber\\
			      &= \left\{\begin{array}{ll}
				      0 & \text{if\ }\hat{x}<0\\
				      \lp\lambda e^{-\lambda\hat{x}}\rp\frac{e^{-\lambda t}f_S(t)}{\mathbb{P}(S<L)} & \text{if\ }\hat{x}>0
			      \end{array}\right.\nonumber\\
			      &= \left\{\begin{array}{ll}
				      0 & \text{if\ }\hat{x}<0\\
				      h(\hat{x})g(t) & \text{if\ }\hat{x}>0
                      \end{array}\right..
\end{align}
Moreover, $\hat{X_i}$ is exponential with rate $\lambda$ since
\begin{align}
	\label{eq:eq_ch3_hat_x}
	\Prob\lp\hat{X}_j>t\rp	&= \Prob\lp L_j>t+S_j|L_j>S_j\rp\nn
				&=\frac{\Prob\lp L_j>t+S_j\rp}{\Prob(L_j>S_j)}\nn
				&= \frac{1}{\Prob(L_j>S_j)}\lp\int_0^\infty e^{-\lambda(t+s)}f_{S_j}(s)\mathrm{d}s\rp\nn
				&= (1+\lambda\theta)^k\lp \frac{e^{-\lambda t}}{(1+\lambda\theta)^k}\rp\nn
				&= e^{-\lambda t}.
\end{align}
\eqref{eq:eq_ch3_eq12} and \eqref{eq:eq_ch3_hat_x} show that $\hat{X}_j$ and $T_j$ are indeed independent. Given that
$\hat{X}_j$ and $Z_j$ are both independent from $T_j$, then $Y_j$ and $T_j$ are also independent.

Furthermore, since $Y_{j-1}=\hat{X}_{j-1}+Z_{j-1}$, $\hat{X}_j$ is independent from $T_j$ and the interarrival process is i.i.d
and independent from the i.i.d service process, then $\hat{X}_j$  and $Z_j$ are independent of $Y_{j-1}$. This implies that for
any $j\geq1$, $Y_{j-1}$ and $Y_j$ are independent. Moreover, it is clear that the $Z_j$'s have the same distribution. Since the
$\hat{X}_j$'s are exponential with rate $\lambda$ then the $(Y_j)_{j\geq1}$ is an i.i.d
process. Given that $Y_j$ is the interval of time between the receptions of two consecutive successful stream-$\str$ packets, then the
number of successfully received packets in the interval $[0,\tau]$, $R(\tau)$, is a renewal process.
\end{proof}
\end{document}